\documentclass[article,shortnames,nojss]{jss}\usepackage[]{graphicx}\usepackage[]{color}
\makeatletter
\def\maxwidth{ %
  \ifdim\Gin@nat@width>\linewidth
    \linewidth
  \else
    \Gin@nat@width
  \fi
}
\makeatother

\usepackage{Sweave}

\usepackage{thumbpdf,lmodern}
\usepackage{amsfonts,amsmath,amssymb,amsthm}
\usepackage{bbm,booktabs,setspace,longtable}
\usepackage[T1]{fontenc}

\newcommand{\class}[1]{`\code{#1}'}
\newcommand{\fct}[1]{\code{#1()}}

\newcommand{\cF}{\mathcal{F}}

\newcommand{\R}{\mathbb{R}}
\newcommand{\by}{\mathbf{y}}
\newcommand{\bX}{\mathbf{X}}
\newcommand{\ind}{\mathbbm{1}}

\newcommand{\crpssample}{\fct{crps\char`_sample}}
\newcommand{\logssample}{\fct{logs\char`_sample}}
\newcommand{\vssample}{\fct{vs\char`_sample}}
\newcommand{\arms}{\fct{ar\char`_ms}}

\DeclareMathOperator{\CRPS}{CRPS}

\author{Alexander Jordan\\University of Bern
	\And Fabian Kr\"uger\\Heidelberg University
	\And Sebastian Lerch\\Heidelberg Institute for\\Theoretical Studies\\Karlsruhe Institute of\\Technology}
\Plainauthor{Alexander Jordan, Fabian Kr\"uger, Sebastian Lerch}

\title{Evaluating Probabilistic Forecasts with \pkg{scoringRules}}
\Plaintitle{Evaluating Probabilistic Forecasts with scoringRules}
\Shorttitle{Evaluating Probabilistic Forecasts}

\Abstract{
  Probabilistic forecasts in the form of probability distributions over future events have become popular in several fields including meteorology, hydrology, economics, and demography. In typical applications, many alternative statistical models and data sources can be used to produce probabilistic forecasts. Hence, evaluating and selecting among competing methods is an important task. The \pkg{scoringRules} package for \proglang{R} provides functionality for comparative evaluation of probabilistic models based on proper scoring rules, covering a wide range of situations in applied work. This paper discusses implementation and usage details, presents case studies from meteorology and economics, and points to the relevant background literature.
}

\Plainkeywords{comparative evaluation, ensemble forecasts, out-of-sample evaluation, predictive distributions, proper scoring rules, score computation, R}

\Address{
  Alexander Jordan\\
  University of Bern\\
  Institute of Mathematical Statistics and Actuarial Science\\
  Alpeneggstrasse 22\\
  3012 Bern, Switzerland\\
  E-Mail: \email{alexander.jordan@stat.unibe.ch}\\
  
  Fabian Kr\"uger\\
  Alfred-Weber-Institute for Economics\\
  Heidelberg University\\
  Bergheimer Str. 58\\
  69115 Heidelberg, Germany\\
  E-Mail: \email{fabian.krueger@awi.uni-heidelberg.de}\\
  URL: \url{https://sites.google.com/site/fk83research/home}\\
  
  Sebastian Lerch\\
  Heidelberg Institute for Theoretical Studies\\
  HITS gGmbH\\
  Schloss-Wolfsbrunnenweg 35\\
  69118 Heidelberg, Germany\\
  E-Mail: \email{sebastian.lerch@h-its.org}\\
  URL: \url{https://sites.google.com/site/sebastianlerch/}\\
  \emph{and}\\
  Institute for Stochastics\\
  Karlsruhe Institute of Technology\\
}
\IfFileExists{upquote.sty}{\usepackage{upquote}}{}
\begin{document}

\section*{Preface}

This vignette corresponds to an article of the same name which has been conditionally accepted for publication at the Journal of Statistical Software. While the two articles are close to identical at the time of this writing (July 27, 2018), the vignette may be updated to reflect future changes in the package. To cite the package, please use \cite{JordanEtAl2018}. 

\section{Introduction: Forecast evaluation} \label{sec:intro}

Forecasts are generally surrounded by uncertainty, and being able to quantify this uncertainty is key to good decision making. Accordingly, probabilistic forecasts in the form of predictive probability distributions over future quantities or events have become popular over the last decades in various fields including meteorology, climate science, hydrology, seismology, economics, finance, demography and political science. Important examples include the United Nation's probabilistic population forecasts \citep{RafteryEtAl2014}, inflation projections issued by the Bank of England \citep[e.g.,][]{Clements2004}, or the now widespread use of probabilistic ensemble methods in meteorology \citep{GneitingRaftery2005, LeutbecherPalmer2008}. For recent reviews see \citet{GneitingKatzfuss2014} and \citet{Raftery2017}. 

With the proliferation of probabilistic models arises the need for tools to evaluate the appropriateness of models and forecasts in a principled way. Various measures of forecast performance have been developed over the past decades to address this demand. Scoring rules are functions $S(F, y)$ that evaluate the accuracy of a forecast distribution $F$, given that an outcome $y$ was observed. As such, they allow to compare alternative models, a crucial ability given the variety of theories, data sources and statistical specifications available in many situations. Conceptually, scoring rules can be thought of as error measures for distribution functions: While the squared error $\text{SE}(x, y) = (y-x)^2$ measures the performance of a point forecast $x$, a scoring rule $S(F, y)$ measures the performance of a distribution forecast $F$. 

This paper introduces the \proglang{R} \citep{R} software package \pkg{scoringRules} \citep{JordanEtAl2016scoringRules}, which provides functions to compute scoring rules for a variety of distributions $F$ that come up in applied work, and popular choices of $S$. Two main classes of probabilistic forecasts are parametric distributions and distributions that are not known analytically, but are indirectly described through a sample of simulation draws. For example, Bayesian forecasts produced via Markov chain Monte Carlo (MCMC) methods take the latter form. Hence, the \pkg{scoringRules} package provides a general framework for model evaluation that covers both classical (frequentist) and Bayesian forecasting methods. 

The \pkg{scoringRules} package aims to be a comprehensive library for computing scoring rules. We offer implementations of several known (but not routinely applied) formulas, and implement some closed-form expressions that were previously unavailable. Whenever more than one implementation variant exists, we offer statistically principled default choices. The package contains the continuous ranked probability score ($\CRPS$) and the logarithmic score, as well as the multivariate energy score and variogram score. All of these scoring rules are proper, which means that forecasters have an incentive to state their true belief, see Section~\ref{sec:background}.

It is worth emphasizing that scoring rules are designed for comparative forecast evaluation. That is, one wants to know whether model A or model B provides better forecasts, in terms of a proper scoring rule. Comparative forecast evaluation is of interest either for choosing a specification for future use, or for comparing various scientific approaches. A distinct, complementary issue is to check the suitability of a given model via tools for absolute forecast evaluation. Here, the focus typically lies in diagnostics, e.g., whether the predictive distributions match the observations statistically \citep[see probability integral transform histogram, e.g., in][]{GneitingKatzfuss2014}. To retain focus, the \pkg{scoringRules} package does not cover absolute forecast evaluation.

Comparative forecast evaluation shares key aspects with out-of-sample model comparison. In that sense, \pkg{scoringRules} is broadly related to all software packages which help users determine an appropriate model for the data at hand. Perhaps most fundamentally, the \pkg{stats} \citep{R} package provides the traditional Akaike and Bayes information criteria to select among linear models in in-sample evaluation. The packages \pkg{caret} \citep{caret} and \pkg{forecast} \citep{HyndmanKhandakar2008} provide cross-validation tools suitable for cross-sectional and time series data, respectively. The \pkg{loo} \citep{loo} package implements recent proposals to select among Bayesian models. In contrast to existing software, a key novelty of the \pkg{scoringRules} package is its extensive coverage of the $\CRPS$. This scoring rule is attractive for both practical and theoretical reasons \citep{GneitingRaftery2007,KruegerEtAl2016}, yet more widespread use has been hampered by computational challenges. In providing analytical formulas and efficient numerical implementations, we hope to enable convenient use of the $\CRPS$ in applied work.

To the best of our knowledge, \pkg{scoringRules} is the first \proglang{R} package designed as a library for computing proper scoring rules for many types of forecast distributions. However, a number of existing \proglang{R} packages include scoring rule computations for more specific empirical situations: The \pkg{ensembleBMA} \citep{ensembleBMA} and \pkg{ensembleMOS} \citep{ensembleMOS} packages contain formulas for the CRPS of a small subset of the distributions listed in Table \ref{tab:parametric-families} which are relevant for post-processing ensemble weather forecasts \citep{FraleyEtAl2011}, and can only be applied to specific data structures utilized in the packages. The \pkg{surveillance} \citep{MeyerEtAl2017} package provides functions to compute the logarithmic score and other scoring rules for count data models in epidemiology. By contrast, the distributions contained in \pkg{scoringRules} are relevant in applications across disciplines and the score functions are generally applicable. The \pkg{scoring} \citep{MerkleSteyvers2013} package focuses on discrete (categorical) outcomes, for which it offers a large number of proper scoring rules. It is thus complementary to \pkg{scoringRules} which supports a wide range of forecast distributions while focusing on a smaller number of scoring rules. Furthermore, the \pkg{verification} \citep{verificationRpackage} and \pkg{SpecsVerification} \citep{SpecsVerificationR} packages contain implementations of the CRPS for simulated forecast distributions. Our contribution in that domain is twofold: First, we offer efficient implementations to deal with predictive distributions given as large samples. MCMC methods are popular across the disciplines, and many sophisticated \proglang{R} implementations are available \citep[e.g.,][]{Kastner2016,CarpenterEtAl2017}. Second, we include various implementation options, and propose principled default settings based on recent research \citep{KruegerEtAl2016}. 

For programming languages other than \proglang{R}, implementations of proper scoring rules are sparse, and generally cover a much narrower score computation functionality than the \pkg{scoringRules} package. The \pkg{properscoring} \citep{properscoringPython} package for the \proglang{Python} \citep{Python} language provides implementations of the CRPS for Gaussian distributions and for forecast distributions given by a discrete sample. Several institutionally supported software packages include tools to compute scoring rules, but typically require input in specific data formats and are tailored towards operational use at meteorological institutions. The \proglang{Model Evaluation Tools} \citep{MET} software provides code to compute the CRPS based on a sample from the forecast distribution. However, note that a Gaussian approximation is applied which can be problematic if the underlying distribution is not Gaussian, see \citet{KruegerEtAl2016}. The \proglang{Ensemble Verification System} \citep{EVS} also provides an implementation of the CRPS for discrete samples. For a general overview of software for forecast evaluation in meteorology, see \citet{Pocernich2012}.

The remainder of this paper is organized as follows. Section~\ref{sec:background} provides some theoretical background on scoring rules, and introduces the logarithmic score and the continuous ranked probability score. Section \ref{sec:usage} gives an overview of the score computation functionality in the \pkg{scoringRules} package and presents the implementation of univariate proper scoring rules. In Section~\ref{sec:examples}, we give usage examples by application in case studies. In a meteorological example of accumulated precipitation forecasts, we compare ensemble system output from numerical weather prediction models to parametric forecast distributions from statistical post-processing models. A second case study shows how using analytical information of a Bayesian time series model for the growth rate of the US economy's gross domestic product (GDP) can help in evaluating the model's simulation output. Definitions and details on the use of multivariate scoring rules are provided in Section \ref{sec:multiv}. The paper closes with a discussion in Section~\ref{sec:discussion}. Two appendices contain various closed-form expressions for the CRPS, as well as details on evaluating multivariate forecast distributions.

\section{Theoretical background}\label{sec:background}

Probabilistic forecasts usually fit one of two categories, parametric distributions or simulated samples. The former type is represented by its cumulative distribution function (CDF) or its probability density function (PDF), whereas the latter is often used if the predictive distribution is not available analytically. Here, we give a brief overview of the theoretical background for comparative forecast evaluation in both cases.

\subsection{Proper scoring rules}

Let $\Omega$ denote the set of possible values of the quantity of interest, $Y$, and let $\cF$ denote a convex class of probability distributions on $\Omega$. A scoring rule is a function
\[
S: \cF \times \Omega \longrightarrow \R \cup \{ \infty \} 
\]
that assigns numerical values to pairs of forecasts $F \in \cF$ and observations $y \in \Omega$. For now, we restrict our attention to univariate observations and set $\Omega = \R$ or subsets thereof, and identify probabilistic forecasts $F$ with the associated CDF $F$ or PDF $f$. In Section~\ref{sec:multiv}, we will consider multivariate scoring rules for which $\Omega = \R^d$.

We consider scoring rules to be negatively oriented, such that a lower score indicates a better forecast. For a proper scoring rule, the expected score is optimized if the true distribution of the observation is issued as a forecast, i.e., if
\[
\E_{Y \sim G} \, S(G,Y) \leq \E_{Y \sim G} \, S(F,Y) 
\]
for all $F,G\in\cF$. A scoring rule is further called strictly proper if equality holds only if $F = G$. Using a proper scoring rule is critical for comparative evaluation, i.e., the ranking of forecasts. In practice, the lowest average score over multiple forecast cases among competing forecasters indicates the best predictive performance, and in this setup, proper scoring rules compel forecasters to truthfully report what they think is the true distribution. See \citet{GneitingRaftery2007} for a more detailed review of the mathematical properties.

\begin{table}[p]
	\begin{center}
		\renewcommand{\arraystretch}{1.2}
		\begin{tabular}{@{}llccl@{}}
			\toprule
			\textbf{Distribution} & \textbf{Family arg.}       & CRPS       & LogS       & \textbf{Additional parameters} \\
			\midrule
			\multicolumn{5}{@{}l@{}}{\textbf{Distributions for variables on the real line}} \\
			Laplace               & \code{"lapl"}    & \checkmark & \checkmark & \\
			logistic              & \code{"logis"}   & \checkmark & \checkmark & \\
			normal                & \code{"norm"}    & \checkmark & \checkmark & \\
			mixture of normals    & \code{"mixnorm"} & \checkmark & \checkmark & \\
			Student's $t$         & \code{"t"}       & \checkmark & \checkmark & \\
			two-piece exponential & \code{"2pexp"}   & \checkmark & \checkmark & \\
			two-piece normal      & \code{"2pnorm"}  & \checkmark & \checkmark & \\
			\multicolumn{5}{@{}l@{}}{\textbf{Distributions for non-negative variables}} \\
			exponential           & \code{"exp"}     & \checkmark & \checkmark & \\
			gamma                 & \code{"gamma"}   & \checkmark & \checkmark & \\
			log-Laplace           & \code{"llapl"}   & \checkmark & \checkmark & \\
			log-logistic          & \code{"llogis"}  & \checkmark & \checkmark & \\
			log-normal            & \code{"lnorm"}   & \checkmark & \checkmark & \\
			\multicolumn{5}{@{}l@{}}{\textbf{Distributions with flexible support and/or point masses}} \\
			beta                  & \code{"beta"}    & \checkmark & \checkmark & limits \\
			uniform               & \code{"unif"}    & \checkmark & \checkmark & limits, point masses \\
			exponential           & \code{"exp2"}    &            & \checkmark & location, scale \\
			& \code{"expM"}    & \checkmark &            & location, scale, point mass \\			                      
			gen. extreme value    & \code{"gev"}     & \checkmark & \checkmark & \\
			gen. Pareto           & \code{"gpd"}     & \checkmark & \checkmark & point mass {\footnotesize(CRPS only)} \\
			logistic              & \code{"tlogis"}  & \checkmark & \checkmark & limits \footnotesize(truncation)\\
			& \code{"clogis"}  & \checkmark &            & limits \footnotesize(censoring) \\
			& \code{"gtclogis"} & \checkmark &           & limits, point masses \\
			normal                & \code{"tnorm"}   & \checkmark & \checkmark & limits \footnotesize(truncation) \\
			& \code{"cnorm"}   & \checkmark &            & limits \footnotesize(censoring) \\
			& \code{"gtcnorm"} & \checkmark &            & limits, point masses \\
			Student's $t$         & \code{"tt"}      & \checkmark & \checkmark & limits \footnotesize(truncation) \\
			& \code{"ct"}      & \checkmark &            & limits \footnotesize(censoring) \\
			& \code{"gtct"}    & \checkmark &            & limits, point masses \\
			\multicolumn{5}{@{}l@{}}{\textbf{Distributions for discrete variables}} \\
			binomial              & \code{"binom"}   & \checkmark & \checkmark & \\
			hypergeometric        & \code{"hyper"}   & \checkmark & \checkmark & \\
			negative binomial     & \code{"nbinom"}  & \checkmark & \checkmark & \\
			Poisson               & \code{"pois"}    & \checkmark & \checkmark & \\
			\bottomrule
		\end{tabular}
	\end{center}
	\caption{List of implemented parametric families for which CRPS and LogS can be computed via \fct{crps} and \fct{logs}. The character string is the corresponding value for the \code{family} argument. The CRPS formulas are given in Appendix~\ref{app:crpsformulas}. \label{tab:parametric-families}}	
\end{table}

Popular examples of proper scoring rules for $\Omega = \R$ include the logarithmic score and the continuous ranked probability score. The logarithmic score \citep[LogS;][]{Good1952} is defined as 
\[
\operatorname{LogS}(F,y) = - \log(f(y)), 
\]
where $F$ admits a PDF $f$, and is a strictly proper scoring rule relative to the class of probability distributions with densities.
The continuous ranked probability score \citep{MathesonWinkler1976} is defined in terms of the predictive CDF $F$ and is given by
\begin{equation}\label{eq:crps}
\operatorname{CRPS}(F,y) = \int_\R (F(z) - \ind\{y \leq z\})^2 \, \mathrm{d} z,
\end{equation}
where $\ind\{y \leq z\}$ denotes the indicator function which is one if $y \leq z$ and zero otherwise. If the first moment of $F$ is finite, the CRPS can be written as 
\[
\operatorname{CRPS}(F,y) = \E_{F} | X_1 - y | - \frac{1}{2} \E_{F,F} | X_1 - X_2 |,
\]
where $X_1$ and $X_2$ are independent random variables with distribution $F$, see \citet{GneitingRaftery2007}. The CRPS is a strictly proper scoring rule for the class of probability distributions with finite first moment. Closed-form expressions of the integral in Equation~\ref{eq:crps} can be obtained for many parametric distributions and allow for exact and efficient computation of the CRPS. They are implemented in the \pkg{scoringRules} package for a range of parametric families, see Table~\ref{tab:parametric-families} for an overview, and are provided in Appendix~\ref{app:crpsformulas}.

\subsection{Model assessment based on simulated forecast distributions}
\label{sec:simd}

In various applications, the forecast distribution of interest $F$ is not available in an analytic form, but only through a simulated sample $X_1, \dots, X_m \sim F$. Examples include Bayesian forecasting applications where the sample is generated by a MCMC algorithm, or ensemble weather forecasting applications where the different sample values are generated by numerical weather prediction models with different model physics and/or initial conditions. In order to compute the value of a proper scoring rule, the simulated sample needs to be converted into an estimated distribution (say, $\hat F_m(z)$) that can be evaluated at any point $z \in \mathbb{R}$. The implementation choices and default settings in the \pkg{scoringRules} package follow the findings of \citet{KruegerEtAl2016} who provide a systematic analysis of probabilistic forecasting based on MCMC output.

For the CRPS, the empirical CDF 
\[
\hat{F}_m(z) = \frac{1}{m} \sum_{i=1}^m \ind\{X_i \leq z\}
\]
is a natural approximation of the predictive CDF. In this case, the CRPS reduces to
\begin{equation}\label{eq:crps-kernel-repr}
\operatorname{CRPS}(\hat F_m,y) = \frac{1}{m} \sum_{i=1}^m |X_i - y| - \frac{1}{2 m^2} \sum_{i=1}^m \sum_{j=1}^m |X_i - X_j|
\end{equation}
which allows one to compute the CRPS directly from the simulated sample, see \citet{GrimitEtAl2006}. Implementations of Equation~\ref{eq:crps-kernel-repr} are rather inefficient with computational complexity $\mathcal{O}(m^2)$, and can be improved upon with representations using the order statistics $X_{(1)}, \ldots, X_{(m)}$, i.e., the sorted simulated sample, thus achieving an average $\mathcal{O}(m\log m)$ performance. In the \pkg{scoringRules} package, we use an algebraically equivalent representation of the CRPS based on the generalized quantile function \citep{LaioTamea2007}, leading to
\begin{equation}\label{eq:crps-computation}
\mathrm{CRPS}(\hat F_m, y) = \frac{2}{m^2} \sum_{i = 1}^m (X_{(i)} - y)\left(m \ind\{y < X_{(i)}\} - i + \frac{1}{2}\right),
\end{equation}
which \citet{Murphy1970} reported in the context of the precursory, discrete version of the CRPS. We refer to \citet{Jordan2016} for details.

In contrast to the CRPS, the computation of LogS requires a predictive density. An estimator can be obtained with classical nonparametric kernel density estimation \citep[KDE, e.g.][]{Silverman1986}. However, this estimator is valid only under stringent theoretical assumptions and can be fragile in practice: If the outcome falls into the tails of the simulated forecast distribution, the estimated score may be highly sensitive to the choice of the bandwidth tuning parameter. In an MCMC context, a mixture-of-parameters estimator that utilizes a simulated sample of parameter draws rather than draws from the posterior predictive distribution is a better and often much more efficient choice, see \citet{KruegerEtAl2016}. This mixture-of-parameters estimator is specific to the model being used, in that one needs to know the analytic form of the forecast distribution conditional on the parameter draws. If such knowledge is available, the mixture-of-parameters estimator can be implemented using functionality for parametric forecast distributions. We provide an example in Section~\ref{sec:example-econ}. This example features a conditionally Gaussian forecast distribution, a typical case in applications.

\section{Package design and functionality}\label{sec:usage}


The main functionality of \pkg{scoringRules} is the computation of scores. The essential functions for this purpose follow the naming convention \fct{[score]\char`_[suffix]}, where the two maturest choices for \code{[score]} are \code{crps} and \code{logs}. Regarding the \code{[suffix]}, we aim for analogy to the usual d/p/q/r functions for parametric families of distributions, both in terms of naming convention and usage details, e.g.,
\begin{Code}
crps_norm(y, mean = 0, sd = 1, location = mean, scale = sd)
\end{Code}
Based on these computation functions, package developers may write S3 methods that hook into the respective S3 generic functions, currently limited to
\begin{Code}
crps(y, ...)
logs(y, ...)
\end{Code}
As the implementation of additional scoring rules matures, this list will be extended. We reserve methods for the class \class{numeric}, e.g.,
\begin{Code}
crps.numeric(y, family, ...)
\end{Code}
which are pedantic wrappers for the corresponding \fct{[score]\char`_[family]} functions, but with meaningful error messages, making the \class{numeric} class methods more suitable for interactive use as opposed to numerical score optimization or package integration. Table~\ref{tab:parametric-families} gives a list of the implemented parametric families, as does the \class{numeric} class method documentation for the respective score, e.g., see \code{?crps.numeric}.

Echoing the distinction in Section~\ref{sec:background} between parametric and empirical predictive distributions, we note that computation functions following the naming scheme \fct{[score]\char`_sample}, see Sections~\ref{sec:empirical} and \ref{sec:multiv}, have a special status and cannot be called via the \class{numeric} class method.

\subsection{Parametric predictive distributions}

When the predictive distribution comes from a parametric family, we have two options to perform the score computation and get the resulting vector of score values, i.e., via the score generics or the computation function. For a Gaussian distribution:
\begin{Schunk}
\begin{Sinput}
R> library("scoringRules")
R> obs <- rnorm(10)
R> crps(obs, "norm", mean = c(1:10), sd = c(1:10))
\end{Sinput}
\begin{Soutput}
 [1] 0.288 1.625 1.570 2.003 2.744 3.688 3.270 4.884 4.162 6.067
\end{Soutput}
\begin{Sinput}
R> crps_norm(obs, mean = c(1:10), sd = c(1:10))
\end{Sinput}
\begin{Soutput}
 [1] 0.288 1.625 1.570 2.003 2.744 3.688 3.270 4.884 4.162 6.067
\end{Soutput}
\end{Schunk}
The results are identical, except when the much stricter input checks of the \class{numeric} class method are violated. This helps in detecting manual errors during interactive use, or in debugging automated model fitting and evaluation. Other restrictions in the \class{numeric} class method include the necessity to pass input to all arguments, i.e., default values in the computation functions are ignored, and that all numerical arguments should be vectors of the same length, with the exception that vectors of length one will be recycled. In contrast, the computation functions aim to closely obey standard \proglang{R} conventions.

In package development, we expect predominant use of the computation functions, where developers will handle regularity checks themselves. As a trivial example, we define functions that only depend on the argument \code{y} and compute scores for a fixed predictive gamma distribution:
\begin{Schunk}
\begin{Sinput}
R> crps_y <- function(y) crps_gamma(y, shape = 2, scale = 1.5)
R> logs_y <- function(y) logs_gamma(y, shape = 2, scale = 1.5)
\end{Sinput}
\end{Schunk}
\begin{figure}	
\begin{Schunk}

{\centering \includegraphics[width=\linewidth]{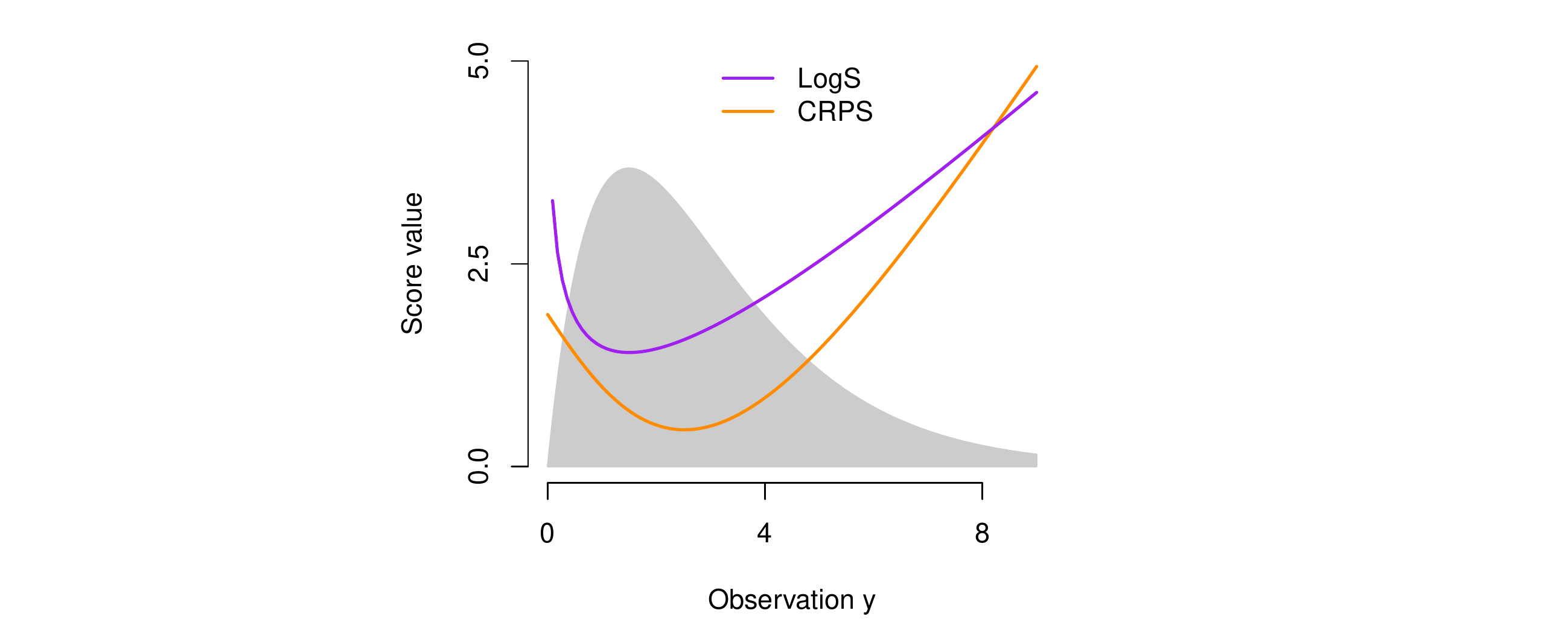} 

}

\end{Schunk}
\caption{Score values for a gamma distribution with \code{shape = 2} and \code{scale = 1.5}, dependent on the observation (functions \fct{crps\_y} and \fct{logs\_y} defined in the text). A scaled version of the predictive density is shown in gray.\label{fig:score-illustration}}
\end{figure}
In Figure~\ref{fig:score-illustration} these functions are used to illustrate the dependence between the score value and the observation in an example of a gamma distribution as forecast. The logarithmic score rapidly increases at the right-sided limit of zero, and the minimum score value is attained if the observation equals the predictive distribution's mode. By contrast, the CRPS is more symmetric around the minimum that is attained at the median value of the forecast distribution, particularly, it increases more slowly as the observation approaches zero.

\newpage

\subsection{Simulated predictive distributions}\label{sec:empirical}

Often forecast distributions can only be given as simulated samples, e.g., ensemble systems in weather prediction (Section~\ref{sec:example-pp}) or MCMC output in econometrics (Section~\ref{sec:example-econ}). We provide functions for both univariate and multivariate samples. The latter are discussed in Section~\ref{sec:multiv}, whereas the former are presented here:
\begin{Code}
crps_sample(y, dat, method = "edf", w = NULL, bw = NULL, 
  num_int = FALSE, show_messages = TRUE)
logs_sample(y, dat, bw = NULL, show_messages = FALSE)
\end{Code}
The input for \code{y} is a vector of observations, and the input for \code{dat} is a matrix with the number of rows matching the length of \code{y} and each row comprising one simulated sample, e.g., for a Gaussian predictive distribution:
\begin{Schunk}
\begin{Sinput}
R> obs_n <- c(0, 1, 2)
R> sample_nm <- matrix(rnorm(3e4, mean = 2, sd = 3), nrow = 3)
R> crps_sample(obs_n, dat = sample_nm)
\end{Sinput}
\begin{Soutput}
[1] 1.216 0.833 0.710
\end{Soutput}
\begin{Sinput}
R> logs_sample(obs_n, dat = sample_nm)
\end{Sinput}
\begin{Soutput}
[1] 2.29 2.10 2.04
\end{Soutput}
\end{Schunk}
When \code{y} has length one then \code{dat} may also be a vector. Random sampling from the forecast distribution can be seen as an option to approximate the values of the proper scoring rules. To empirically assess the quality of this approximation and to illustrate the use of the score functions, consider the following Gaussian toy example, where we examine the performance of forecasts given as samples of size up to $5\,000$. The approximation experiment is independently repeated $500$ times:
\begin{Schunk}
\begin{Sinput}
R> R <- 500
R> M <- 5e3
R> mgrid <- exp(seq(log(50), log(M), length.out = 51))
R> crps_approx <- matrix(NA, nrow = R, ncol = length(mgrid))
R> logs_approx <- matrix(NA, nrow = R, ncol = length(mgrid))
R> obs_1 <- 2
R> for (r in 1:R) {
+    sample_M <- rnorm(M, mean = 2, sd = 3)
+    for (i in seq_along(mgrid)) {
+      m <- mgrid[i]
+      crps_approx[r, i] <- crps_sample(obs_1, dat = sample_M[1:m])
+      logs_approx[r, i] <- logs_sample(obs_1, dat = sample_M[1:m])
+    }
+  }
\end{Sinput}
\end{Schunk}
\begin{figure}
\begin{Schunk}

{\centering \includegraphics[width=\linewidth]{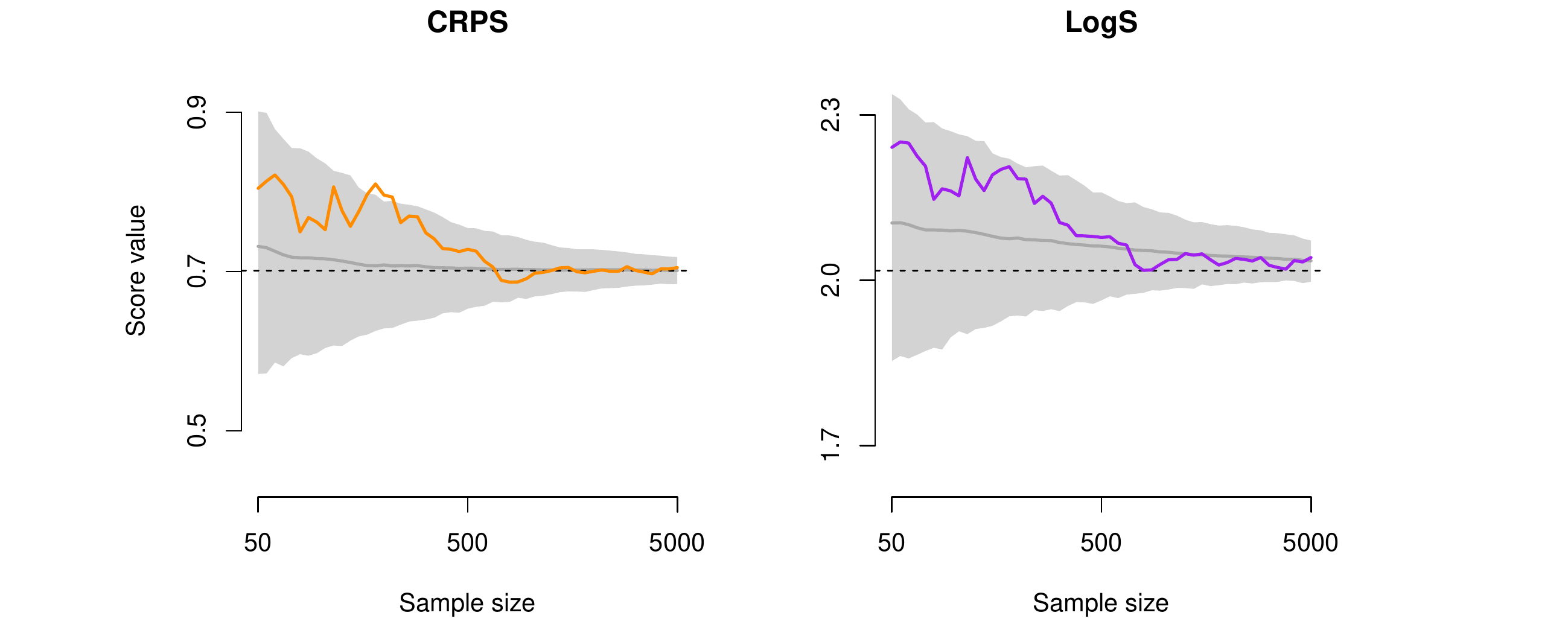} 

}

\end{Schunk}
\caption{\label{fig:score_approx} The scores of a Gaussian forecast distribution with mean 2 and standard deviation 3 when a value of 0 is observed, estimated from an independent sample from the predictive distribution, and shown as a function of the size of the (expanding) sample. The horizontal dashed line represents the analytically calculated score. The gray area indicates empirical 90\% confidence intervals for each sample size computed from 500 independent replications of the simulation experiment, and the gray line shows the corresponding mean value over all repetitions.}
\end{figure}
In this case, the true CRPS and LogS values can be computed using the \fct{crps} and \fct{logs} functions. Figure~\ref{fig:score_approx} graphically illustrates how the scores based on sampling approximations become more accurate as the sample size increases.

The \code{method} argument controls which approximation method is used in \crpssample{}, with possible choices given by \code{"edf"} (empirical distribution function) and \code{"kde"} (kernel density estimation). The default choice \code{"edf"} corresponds to computing the approximation from Equation~\ref{eq:crps-kernel-repr}, implemented as in Equation~\ref{eq:crps-computation}. A vector or matrix of weights, matching the input for \code{dat}, can be passed to the argument \code{w} to compute the CRPS, for any distribution with a finite number of outcomes.

For kernel density estimation, i.e., the default in \logssample{} and the corresponding \code{method} in \crpssample{}, we use a Gaussian kernel to estimate the predictive distribution. Kernel density estimation is an unusual choice in the case of the CRPS, but it is the only implemented option for evaluating the LogS of a simulated sample. In particular, the empirical distribution function is not applicable to LogS because an estimated density is required. The \code{bw} argument allows one to manually select a bandwidth parameter for kernel density estimation; by default, the \fct{bw.nrd} function from the \pkg{stats} \citep{R} package is employed.

\section{Usage examples}\label{sec:examples}

\subsection{Probabilistic weather forecasting via ensemble post-processing}\label{sec:example-pp}

In numerical weather prediction (NWP), physical processes in the atmosphere are modeled through systems of partial differential equations that are solved numerically on three-dimensional grids. To account for major sources of uncertainty, weather forecasts are typically obtained from multiple runs of NWP models with varying initial conditions and model physics resulting in a set of deterministic predictions, called the forecast ensemble. While ensemble predictions are an important step from deterministic to probabilistic forecasts, they tend to be biased and underdispersive (such that, empirically, the actual observation falls outside the range of the ensemble too frequently). Hence, ensembles require some form of statistical post-processing. Over the past decade, a variety of approaches to statistical post-processing has been proposed, including non-homogeneous regression \citep{GneitingEtAl2005} and Bayesian model averaging \citep{RafteryEtAl2005}. 

Here we illustrate how to evaluate post-processed ensemble forecasts of precipitation, based on data and methods from the \pkg{crch} \citep{MessnerEtAl2016} package for \proglang{R}. We model the conditional distribution of precipitation accumulation, $Y \geq 0$, given the ensemble forecasts $X_1,\dots,X_m$ using censored non-homogeneous regression models of the form
\begin{eqnarray}
\Prob(Y = 0|X_1,\dots,X_m) &=& F_{\theta}(0) \label{eq:pproc1}, \\
\Prob(Y \le y|X_1,\dots,X_m) &=& F_{\theta}(y),~\text{for}~y >0, \label{eq:pproc2}
\end{eqnarray}
where $F_\theta$ is the CDF of a continuous parametric distribution with parameters $\theta$. Equations~\ref{eq:pproc1} and \ref{eq:pproc2} specify a mixed discrete-continuous forecast distribution for precipitation: There is a positive probability of observing no precipitation at all ($Y = 0$), however, if $Y > 0$, it can take many possible values $y$. In order to incorporate information from the raw forecast ensemble, we let $\theta$ be a function of $X_1,\dots,X_m$, i.e., we use features of the raw ensemble to determine the parameters of the forecast distribution. Specifically, we consider different location-scale families $F_\theta$ and model the location parameter $\mu$ as a linear function of the ensemble mean $\bar X = \frac{1}{m} \sum_{i = 1}^{m} X_i$,
\[
\mu = a_0 + a_1 \bar X,
\]  
and the scale parameter $\sigma$ as a linear function of the logarithm of the standard deviation $s$ of the ensemble,
\[
\log(\sigma) = b_0 + b_1\log\left( s \right).
\]
A logarithmic link function is used to ensure positivity of the scale parameter. The coefficients $a_0, a_1, b_0, b_1$ can be estimated using maximum likelihood approaches implemented in the \pkg{crch} package. The choice of a suitable parametric family $F_\theta$ is not obvious. Following \citet{MessnerEtAl2016}, we thus consider three alternative choices: the logistic, Gaussian, and Student's $t$ distributions. For details and further alternatives, see, e.g., \citet{MessnerEtAl2014, Scheuerer2014} and \citet{ScheuererHamill2015}.

The \pkg{crch} package contains a data set \code{RainIbk} of precipitation for Innsbruck, Austria. It comprises ensemble forecasts \code{rainfc.1}, \ldots, \code{rainfc.11} and observations \code{rain} for 4971 cases from January 2000 to September 2013. The precipitation amounts are accumulated over three days, and the corresponding 11 member ensemble consists of accumulated precipitation amount forecasts between five and eight days ahead. Following \citet{MessnerEtAl2016} we model the square root of precipitation amounts, and omit forecast cases where the ensemble has a standard deviation of zero. From \citet{MessnerEtAl2016}:
\begin{Schunk}
\begin{Sinput}
R> library("crch")
R> data("RainIbk", package = "crch")
R> RainIbk <- sqrt(RainIbk)
R> ensfc <- RainIbk[, grep('^rainfc', names(RainIbk))]
R> RainIbk$ensmean <- apply(ensfc, 1, mean)
R> RainIbk$enssd <- apply(ensfc, 1, sd)
R> RainIbk <- subset(RainIbk, enssd > 0)
\end{Sinput}
\end{Schunk}
We split the data into a training set until November 2004, and an out-of-sample evaluation period (or test sample) from January 2005.
\begin{Schunk}
\begin{Sinput}
R> data_train <- subset(RainIbk, as.Date(rownames(RainIbk)) <= "2004-11-30")
R> data_eval <- subset(RainIbk, as.Date(rownames(RainIbk)) >= "2005-01-01")
\end{Sinput}
\end{Schunk}
Then, we estimate the censored regression models that are based on the logistic, Student's $t$, and Gaussian distributions, and produce the parameters of the forecast distributions for the evaluation period using built-in functionality of the \pkg{crch} package. We only show the code for the Gaussian model since it can be adapted straightforwardly for the logistic and Student's $t$ models.
\begin{Schunk}
\begin{Sinput}
R> CRCHgauss <- crch(rain ~ ensmean | log(enssd), data_train,
+    dist = "gaussian", left = 0)
R> gauss_mu <- predict(CRCHgauss, data_eval, type = "location")
R> gauss_sc <- predict(CRCHgauss, data_eval, type = "scale")
\end{Sinput}
\end{Schunk}

The raw ensemble of forecasts is a natural benchmark for comparison since interest commonly lies in quantifying the gains in forecast accuracy that result from post-processing:
\begin{Schunk}
\begin{Sinput}
R> ens_fc <- data_eval[, grep('^rainfc', names(RainIbk))]
\end{Sinput}
\end{Schunk}
\begin{figure}
\begin{Schunk}

{\centering \includegraphics[width=\linewidth]{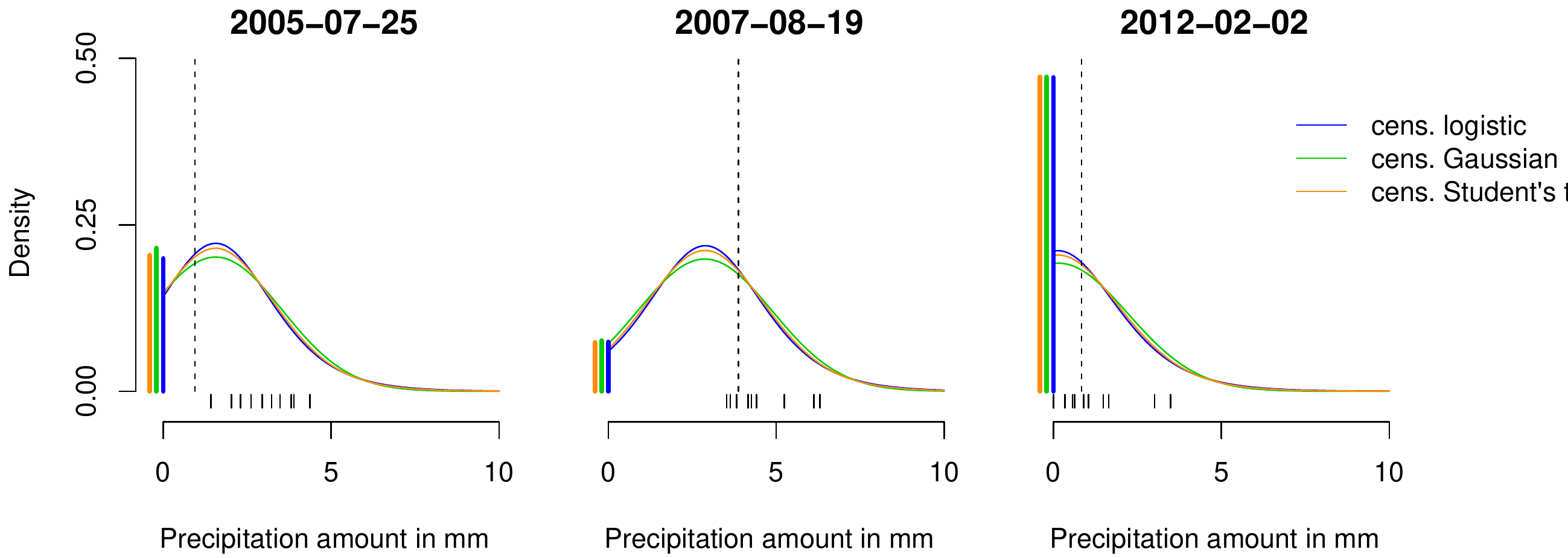} 

}

\end{Schunk}
\caption{Illustration of the forecast distributions of the censored regression models for three illustrative 3-day accumulation periods (plot title indicates end of period). The predicted probabilities of zero precipitation are shown as solid thick vertical lines at 0, and the colored thin lines indicate the upper tail on the positive half axis of the forecast densities $f_\theta$, c.f., Equations~\ref{eq:pproc1} and \ref{eq:pproc2}. The raw ensemble forecasts are shown as short line segments at the bottom, and the realizing observation is indicated by the long dashed line. \label{fig:postprocplot}}
\end{figure}
Figure~\ref{fig:postprocplot} shows the models' forecast distributions in three illustrative cases. To evaluate forecast performance in the entire out of sample period, we use the function \fct{crps} for the model outputs and the function \crpssample{} to compute the CRPS of the raw ensemble. Note that we have to turn \code{ens_fc} into an object of class \class{matrix} manually.
\begin{Schunk}
\begin{Sinput}
R> obs <- data_eval$rain
R> gauss_crps <- crps(obs, family = "cnorm", location = gauss_mu, 
+    scale = gauss_sc, lower = 0, upper = Inf)
R> ens_crps <- crps_sample(obs, dat = as.matrix(ens_fc))
\end{Sinput}
\end{Schunk}

The mean CRPS values indicate that all post-processing models substantially improve upon the raw ensemble forecasts. There are only small differences between the censored regression models, with the models based on the logistic and Student's $t$ distributions slightly outperforming the model based on a normal distribution.
\begin{Schunk}
\begin{Sinput}
R> scores <- data.frame(CRCHlogis = logis_crps, CRCHgauss = gauss_crps,
+    CRCHstud = stud_crps, Ensemble = ens_crps)
R> sapply(scores, mean)
\end{Sinput}
\begin{Soutput}
CRCHlogis CRCHgauss  CRCHstud  Ensemble 
    0.875     0.876     0.875     1.321 
\end{Soutput}
\end{Schunk}

\subsection[Bayesian forecasts of US GDP growth rate]{Bayesian forecasts of {US~GDP} growth rate}\label{sec:example-econ}

We next present a representative example from economics, where the predictive distribution is given as a simulated sample. \citet{Hamilton1989} first proposed the Markov switching autoregressive model to allow regime-dependent modeling, i.e., to capture different recurring time-series characteristics. We consider the following simple variant of the model:
\begin{eqnarray*}
Y_t &=& c_0 + c_1~Y_{t-1} + \varepsilon_t,\\
\varepsilon_t &\sim&\mathcal{N}(0, \sigma^2_{s_t}),
\end{eqnarray*}
where $Y_t$ is the observed GDP growth rate of quarter $t$, and $s_t \in \{1, 2\}$ is a latent state variable that represents two regimes in the residual variance $\sigma^2_{s_t}$. Since $s_t$ evolves according to a first-order Markov chain, the specification allows for the possibility that periods of high (or low) volatility cluster over time. The latter is a salient feature of the US GDP growth rate: For example, the series was much more volatile in the 1970s than in the 1990s. The model is estimated using Bayesian Markov chain Monte Carlo methods \citep[e.g.,]{Fruhwirth2006}. Our implementation closely follows \citet[Section 5]{KruegerEtAl2016}, and uses the \arms{} function, and the data set \code{gdp}, included in the \pkg{scoringRules} package.

The data set \code{gdp} comprises US GDP growth observations \code{val}, and the corresponding quarters \code{dt} and vintages \code{vint}. Measuring economic variables is challenging, hence records tend to be revised and each quarter has its own time-series, or vintage, of past observations. As a result, the data set for 271 quarters from 1947 to 2014 contains 33616 records.

We split the data into a training sample of observations containing the data before 2014's first quarter, and an evaluation period containing only the four quarters of 2014, using the most recent vintage in both cases:
\begin{Schunk}
\begin{Sinput}
R> data("gdp", package = "scoringRules")
R> data_train <- subset(gdp, vint == "2014Q1")
R> data_eval <- subset(gdp, vint == "2015Q1" & grepl("2014", dt))
\end{Sinput}
\end{Schunk}
As is typical for MCMC-based analysis, the model's forecast distribution $F_0$ is not available as an analytical formula, but must be approximated in some way. Following \citet{KruegerEtAl2016}, a generic MCMC algorithm to generate samples of the parameter vector $\theta$ and sample from the posterior predictive distribution proceeds as follows:
\begin{itemize}
	\item fix $\theta_0\in\Theta$
	\item for $i = 1,\dots,m$, 
	\begin{itemize}
		\item draw $\theta_i \sim \mathcal{K}(\cdot|\theta_{i-1})$, where $\mathcal{K}$ is a transition kernel that is specific to the model under use
		\item draw $X_i\sim F_c(\cdot|\theta_i)$, where $F_c$ denotes the conditional distribution given the parameter values.
	\end{itemize}
\end{itemize}
We use the function \arms{} to fit the model and produce forecasts for the four quarters of 2014 based on the information available at the end of year 2013, i.e., a single prediction case where the forecast horizon extends from one to four quarters ahead. Here, the conditional distribution $F_c$ is Gaussian, and we run the chain for 20\,000 iterations.
\begin{Schunk}
\begin{Sinput}
R> h <- 4
R> m <- 20000
R> fc_params <- ar_ms(data_train$val, forecast_periods = h, n_rep = m)
\end{Sinput}
\end{Schunk}
This function call yields a simulated sample corresponding to $\{\theta_1, \ldots, \theta_m\}$, where we obtain matrices of parameters for the mean and standard deviation. We transpose these matrices to have the rows correspond to the observations, and columns represent the position in the Markov chain:
\begin{Schunk}
\begin{Sinput}
R> mu <- t(fc_params$fcMeans)
R> Sd <- t(fc_params$fcSds)
\end{Sinput}
\end{Schunk}
Next, we draw the sample of possible observations corresponding to $\{X_1, \ldots, X_m\}$ conditional on the Gaussian assumption and the available parameter information:
\begin{Schunk}
\begin{Sinput}
R> X <- matrix(rnorm(h * m, mean = mu, sd = Sd), nrow = h, ncol = m)
\end{Sinput}
\end{Schunk}
We consider two competing estimators of the posterior predictive distribution $F_0$. The mixture-of-parameters estimator (MPE)
\begin{equation}
\hat{F}_m^{\text{MP}}(z) = \frac{1}{m} \sum_{i=1}^{m} F_c(z|\theta_i),\label{eqn:mix}
\end{equation}
builds on the simulated parameter values by mixing a series of Gaussian distributions uniformly, whereas the empirical CDF based approximation
\[
\hat{F}_m^{\text{ECDF}}(z) = \frac{1}{m} \sum_{i=1}^m \ind\{X_i \leq z\}
\]
utilizes the simulated sample from the conditional distribution given the parameter values, instead of building on the simulated parameter values directly. A standard choice for a smoother approximation is to replace the indicator function with a Gaussian kernel, as in the \logssample{} function.

\begin{figure}
\begin{Schunk}

{\centering \includegraphics[width=\linewidth]{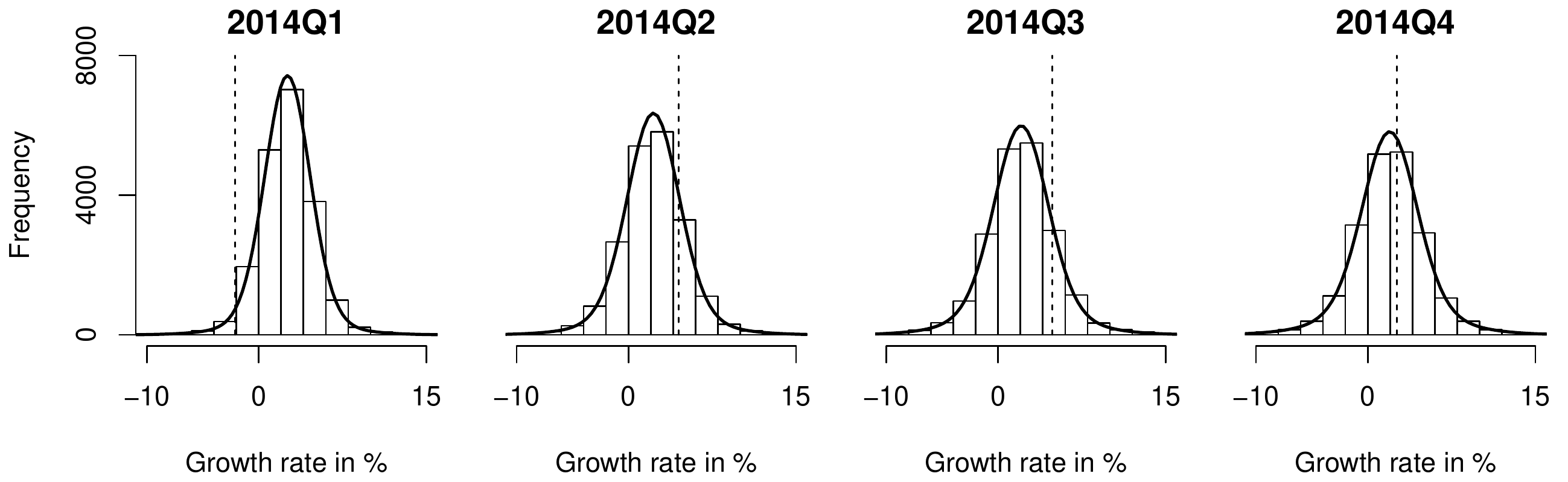} 

}

\end{Schunk}
\caption{Forecast distributions for the growth rate of US GDP. The forecasts stem from a Bayesian time series model, as detailed in \citet[Section 5]{KruegerEtAl2016}. Histograms summarize simulated forecast draws at each date. Mixture-of-normals approximation to distribution shown in black; realizing observations shown by dashed line.	\label{fig:mcmcplot}}	
\end{figure}

The two alternative estimators are illustrated in Figure~\ref{fig:mcmcplot}: For each date, the histogram represents a simulated sample from the model's forecast distribution, and the black line indicates the mixture-of-parameters estimator. We can observe a distinct decrease in the forecast's certainty as the forecast horizon increases from one to four quarters ahead.

Finally, we evaluate CRPS and LogS for the approximated forecast distributions described above. The mixture-of-parameters estimator $\hat{F}_m^{\text{MP}}$ can be evaluated with the functions \fct{crps} and \fct{logs}, and $\hat{F}_m^{\text{ECDF}}$ can be evaluated with the functions \crpssample{} and \logssample{}:
\begin{Schunk}
\begin{Sinput}
R> obs <- data_eval$val
R> names(obs) <- data_eval$dt
R> w <- matrix(1/m, nrow = h, ncol = m)
R> crps_mpe <- crps(obs, "normal-mixture", m = mu, s = Sd, w = w)
R> logs_mpe <- logs(obs, "normal-mixture", m = mu, s = Sd, w = w)
R> crps_ecdf <- crps_sample(obs, X)
R> logs_kde <- logs_sample(obs, X)
R> print(cbind(crps_mpe, crps_ecdf, logs_mpe, logs_kde))
\end{Sinput}
\begin{Soutput}
       crps_mpe crps_ecdf logs_mpe logs_kde
2014Q1    3.457     3.468     4.01     3.97
2014Q2    1.358     1.362     2.29     2.28
2014Q3    1.700     1.690     2.52     2.54
2014Q4    0.724     0.729     1.96     1.98
\end{Soutput}
\end{Schunk}
The score values are quite similar for both estimators, which seems natural given the large number of $20\,000$ MCMC draws. For the logarithmic score in particular, the MPE should be preferred over the KDE based estimator on theoretical grounds, see \citet{KruegerEtAl2016}.

The algorithm and approximation methods just sketched are not idiosyncratic to our example, but arise whenever a Bayesian model is used for forecasting. For illustrative \proglang{R} implementations of other Bayesian models, see, e.g., the packages \pkg{bayesgarch} \citep{ArdiaHoogerheide2010} and \pkg{stochvol} \citep{Kastner2016}.

\subsection[Parameter estimation with scoring rules]{Parameter estimation with scoring rules}\label{sec:param-estim}

Apart from comparative forecast evaluation, proper scoring rules also provide useful tools for parameter estimation. In the optimum score estimation framework of \citet[Section 9.1]{GneitingRaftery2007}, the parameters of a model's forecast distribution are determined by optimizing the value of a proper scoring rule, on average over a training sample. Optimum score estimation based on the LogS corresponds to classical maximum likelihood estimation. The score functions to compute $\CRPS$ and $\operatorname{LogS}$ for parametric forecast distributions included in \pkg{scoringRules} (see Table~\ref{tab:parametric-families}) thus offer tools for the straightforward implementation of such optimum score estimation approaches. Specifically, the worker functions of the form \fct{[crps]\char`_[family]} and \fct{[logs]\char`_[family]} entail little overhead in terms of input checks and are thus well suited for use in numerical optimization procedures such as \fct{optim}. Furthermore, functions to compute gradients and Hessian matrices of the $\CRPS$ have been implemented for a subset of parametric families, and can be supplied to assist numerical optimizers. Such functions are available for the \code{"logis", "norm"} and \code{"t"} families and truncated and censored versions thereof (\code{"clogis", "tlogis", "cnorm", "tnorm", "ct", "tt"}). The corresponding computation functions follow the naming scheme \fct{[gradcrps]\char`_[family]} and \fct{[hesscrps]\char`_[family]}. However, we emphasize that implementing minimum CRPS or $\operatorname{LogS}$ estimation approaches is possible for all parametric families listed in Table \ref{tab:parametric-families}, even if analytical gradient and Hessian functions are not supplied but are instead approximated numerically by \fct{optim}.

\citet{GebetsbergerEtAl2017} provide a detailed comparison of maximum likelihood and minimum CRPS estimation in the context of non-homogeneous regression models for post-processing ensemble weather forecasts. Here we illustrate the use for minimum CRPS estimation in a simple simulation example. Consider an independent sample $y_1,\dots,y_n$ from a normal distribution with mean $\mu$ and standard deviation $\sigma$. The analytical maximum likelihood estimates $\hat\mu_{\text{ML}}$ and $\hat\sigma_{\text{ML}}$ are given by
\begin{equation*}
\hat\mu_{\text{ML}} = \frac{1}{n} \sum_{i=1}^{n} y_i \quad \text{and} \quad \hat\sigma_{\text{ML}} = \sqrt{\frac{1}{n}\sum_{i=1}^{n} (y_i - \hat\mu_{\text{ML}})^2}.
\end{equation*}

To determine the corresponding estimates by numerically minimizing the CRPS define wrapper functions which compute the mean CRPS and corresponding gradient for a vector of training data \code{y_train} and distribution parameters \code{param}. 
\begin{Schunk}
\begin{Sinput}
R> meancrps <- function(y_train, param) mean(crps_norm(y = y_train,
+    mean = param[1], sd = param[2]))
R> grad_meancrps <- function(y_train, param) apply(gradcrps_norm(y_train,
+    param[1], param[2]), 2, mean)
\end{Sinput}
\end{Schunk}
These functions can then be passed to \fct{optim}, for example, mean and standard deviation of a normal distribution with true values $-1$ and $2$ can be estimated as illustrated in the following. The estimation with sample size 500 is repeated $1\,000$ times.
\begin{Schunk}
\begin{Sinput}
R> R <- 1000
R> n <- 500
R> mu_true <- -1
R> sigma_true <- 2
R> estimates_ml <- matrix(NA, nrow = R, ncol = 2)
R> estimates_crps <- matrix(NA, nrow = R, ncol = 2)  
R> for (r in 1:R) {
+    dat <- rnorm(n, mu_true, sigma_true)
+    estimates_crps[r, ] <- optim(par = c(1, 1), fn = meancrps,
+      gr = grad_meancrps, method = "BFGS", y_train = dat)$par
+    estimates_ml[r, ] <- c(mean(dat), sd(dat) * sqrt((n - 1) / n))
+  }
\end{Sinput}
\end{Schunk}

\begin{figure}
\begin{Schunk}

{\centering \includegraphics[width=\linewidth]{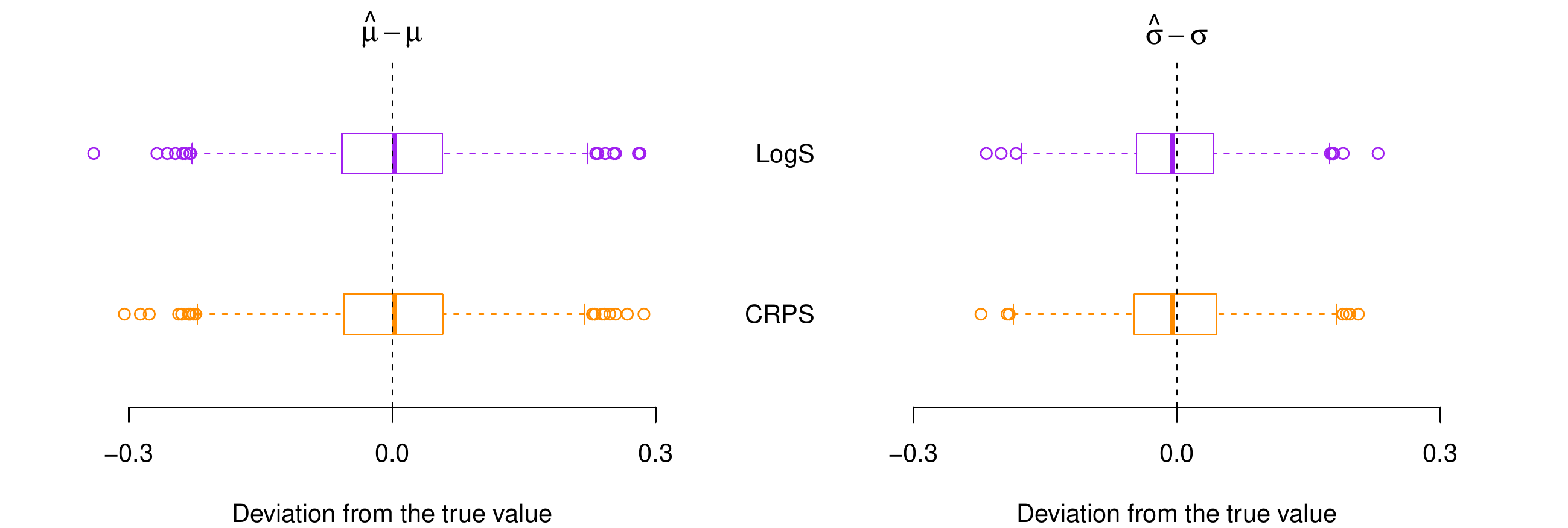} 

}

\end{Schunk}
\caption{Boxplots of deviations from the true parameter values for estimates obtained via minimum CRPS and minimum LogS (i.e., maximum likelihood) estimation based on 1\,000 independent samples of size 500 from a normal distribution with mean $\mu = -1$ and standard deviation $\sigma = 2$. \label{fig:parameter-estimation}}	
\end{figure}

Figure \ref{fig:parameter-estimation} compares minimum CRPS and minimum LogS (i.e., maximum likelihood) parameter estimates. The differences to the true values show very similar distributions and illustrate the consistency of general optimum score estimates \citep[Equation~59]{GneitingRaftery2007}. For the standard deviation parameter $\sigma$, the difference between estimate and true value exhibits slightly less variability for the maximum likelihood method.

\section{Multivariate scoring rules}\label{sec:multiv}

The basic concept of proper scoring rules can be extended to multivariate forecast distributions for which the support $\Omega$ is given by $\R^d, d \in \{2, 3, \ldots \}$. A variety of multivariate proper scoring rules has been proposed in the literature. For example, the univariate LogS allows for a straightforward generalization towards multivariate forecast distributions. However, parametric modeling and forecasting of multivariate observations is challenging, and when sampling is a feasible alternative we encounter the same, even exacerbated, problems in kernel density estimation as for univariate samples. As another example, the univariate CRPS can also be generalized to multivariate forecast distributions, and one such generalization is discussed in this chapter, the energy score. Finding closed form expressions for parametric distributions is even more involved than for the univariate CRPS, but the robustness in the evaluation of sample forecasts is retained. We refer to \citet{GneitingEtAl2008} and \citet{ScheuererHamill2015VS} for a detailed discussion of multivariate proper scoring rules and limit our attention to the case where probabilistic forecasts are given as samples from the forecast distributions.

Let $\by = (y^{(1)},\dots,y^{(d)}) \in\Omega=\R^d$, and let $F$ denote a forecast distribution on $\R^d$ given through $m$ discrete samples $\bX_1,\dots,\bX_m$ from $F$ with $\bX_i = (X_i^{(1)},\dots,X_i^{(d)}) \in \R^d, i=1,\dots,m$. The \pkg{scoringRules} package provides implementations of the energy score \citep[ES;][]{GneitingEtAl2008},
\[
\operatorname{ES}(F,y) = \frac{1}{m}\sum_{i=1}^m \| \bX_i - \by \| - \frac{1}{2m^2} \sum_{i = 1}^m\sum_{j = 1}^m \| \bX_i - \bX_j \|,
\]
where $\|\cdot\|$ denotes the Euclidean norm on $\R^d$, and the variogram score of order $p$ \citep[VS$^p$;][]{ScheuererHamill2015VS},
\[
\operatorname{VS}^p(F,y) = \sum_{i=1}^d\sum_{j=1}^d w_{i,j} \left( \left|y^{(i)} - y^{(j)} \right|^p - \frac{1}{m}\sum_{k=1}^m \left|X_k^{(i)} - X_k^{(j)} \right|^p \right)^2.
\]
In the definition of VS$^p$, $w_{i,j}$ is a non-negative weight that allows one to emphasize or down-weight pairs of component combinations based on subjective expert decisions, and $p$ is the order of the variogram score. Typical choices of $p$ include $0.5$ and $1$.

ES and VS$^p$ are implemented for multivariate forecast distributions given through simulated samples as functions
\begin{Code}
es_sample(y, dat)
vs_sample(y, dat, w = NULL, p = 0.5)
\end{Code}
These functions can only evaluate a single multivariate forecast case and always return a single number to simplify use and documentation, see Appendix \ref{sec:multiv-multiple} for an example on how to use \fct{apply} functions or \code{for} loops to sequentially apply them to multiple forecast cases. The observation input for \code{y} is required to be a vector of length $d$, and the corresponding forecast input for \code{dat} has to be given as a $d\times m$ matrix, the columns of which are the simulated samples $\bX_1,\dots,\bX_m$ from the multivariate forecast distribution. In \vssample{} it is possible to specify a $d\times d$ matrix for \code{w} of non-negative weights as described in the text. The entry in the $i$-th row and $j$-th column of \code{w} corresponds to the weight assigned to the combination of the $i$-th and $j$-th component. If no weights are specified, constant weights with $w_{i,j} = 1$ for all $i,j \in\{ 1,\dots,d\}$ are used. For details and examples on choosing appropriate weights, see \citet{ScheuererHamill2015VS}.

In the following, we give a usage example of the multivariate scoring rules using the results from the economic case study in Section~\ref{sec:example-econ}. Instead of evaluating the forecasts separately for each horizon (as we did before), we now jointly evaluate the forecast performance over the four forecast horizons based on the four-variate simulated sample.
\begin{Schunk}
\begin{Sinput}
R> names(obs) <- NULL
R> es_sample(obs, dat = X)
\end{Sinput}
\begin{Soutput}
[1] 4.13
\end{Soutput}
\begin{Sinput}
R> vs_sample(obs, dat = X)
\end{Sinput}
\begin{Soutput}
[1] 7.05
\end{Soutput}
\end{Schunk}
While this simple example refers to a single forecast case and a single model, a typical empirical analysis would consider the average scores (across several forecast cases) of two or more models.

\section{Summary and discussion}\label{sec:discussion}

The \pkg{scoringRules} package enables computing proper scoring rules for parametric and simulated forecast distributions. The package covers a wide range of situations prevalent in work on modeling and forecasting, and provides generally applicable and numerically efficient implementations based on recent theoretical considerations. 

The main functions of the package -- \fct{crps} and \fct{logs} -- are S3 generics, for which we provide methods \fct{crps.numeric} and \fct{logs.numeric}. The package can be extended naturally by defining S3 methods for classes other than \class{numeric}. For example, consider a fitted model object of class \class{crch}, obtained by the \proglang{R} package of the same name \citep[][]{MessnerEtAl2016}. An object of this class contains a detailed specification of the fitted model's forecast distribution (such as the parametric family of distributions and the values of the fitted parameters). This information could be utilized to write a specific method that computes the $\CRPS$ of a fitted model object. 


The choice of an appropriate proper scoring rule for model evaluation or parameter estimation is a non-trivial task. We have implemented the widely used LogS and $\CRPS$ along with the multivariate $\operatorname{ES}$ and $\operatorname{VS}^p$. Possible future extension of the \pkg{scoringRules} package include the addition of novel proper scoring rules such as the Dawid-Sebastiani score \citep{DawidSebastiani1999} which has been partially implemented. Further, given the availability of appropriate analytical expressions, the list of covered parametric families can be extended as demand arises and time allows.

\section*{Acknowledgments}
The work of Alexander Jordan and Fabian Kr\"uger has been funded by the European Union Seventh Framework Programme under grant agreement 290976. Sebastian Lerch gratefully acknowledges support by Deutsche Forschungsgemeinschaft (DFG) through project C7 (``Statistical postprocessing and stochastic physics for ensemble predictions'') within SFB/TRR 165 ``Waves to Weather''. The authors thank the Klaus Tschira Foundation for infrastructural support at the Heidelberg Institute for Theoretical Studies. Helpful comments by Tilmann Gneiting, Stephan Hemri, Jakob Messner and Achim Zeileis are gratefully acknowledged. We further thank Maximiliane Graeter for contributions to the implementation of the multivariate scoring rules, and two referees for constructive comments on an earlier version of the manuscript.

\bibliography{bibliography}

\begin{appendix}
	\clearpage
	\section{Formulas for the CRPS}\label{app:crpsformulas}
	
	\subsection{Notation}
	
	\begin{longtable}[]{@{}ll@{}}
		\toprule
		Symbol & Name\tabularnewline
		\midrule
		\endhead
		\(\gamma\) & Euler-Mascheroni constant\tabularnewline
		\(\lfloor x \rfloor\) & floor function\tabularnewline
		\(\mathrm{sgn}(x)\) & sign function\tabularnewline
		\(\mathrm{Ei}(x)\) & exponential integral\tabularnewline
		\(\varphi(x)\) & standard Gaussian density function\tabularnewline
		\(\Phi(x)\) & standard Gaussian distribution function\tabularnewline
		\(\Gamma(a)\) & gamma function\tabularnewline
		\(\Gamma_l(a, x)\) & lower incomplete gamma function\tabularnewline
		\(\Gamma_u(a, x)\) & upper incomplete gamma function\tabularnewline
		\(B(a, b)\) & beta function\tabularnewline	
		\(I(a, b, x)\) & regularized incomplete beta function\tabularnewline
		\(I_m(x)\) & modified Bessel function of the first kind\tabularnewline
		\({}_2F_1(a, b; c; x)\) & hypergeometric function\tabularnewline
		\bottomrule
	\end{longtable}
	
	\subsection{Distributions for variables on the real line}
	
	\subsubsection{Laplace distribution}
	The function \fct{crps\char`_lapl} computes the CRPS for the standard distribution, and generalizes via \code{location} parameter $\mu \in \mathbb{R}$ and \code{scale} parameter $\sigma > 0$,
	\[
	\begin{aligned}
	\CRPS(F, y) &= |y| + \exp(-|y|) - \frac{3}{4}, \\
	\CRPS(F_{\mu, \sigma}, y) &= \sigma \CRPS\left(F, \tfrac{y - \mu}{\sigma}\right).
	\end{aligned}
	\]
	The CDFs are given by \(F_{\mu, \sigma}(x) = F\left(\tfrac{x - \mu}{\sigma}\right)\) and
	\[
	F(x) = \begin{cases} \frac{1}{2} \exp(x), & x < 0,\\ 1 - \frac{1}{2} \exp(-x), & x \geq 0. \end{cases}
	\]
	
	\subsubsection{Logistic distribution}
	The function \fct{crps\char`_logis} computes the CRPS for the standard distribution, and generalizes via \code{location} parameter $\mu \in \mathbb{R}$ and \code{scale} parameter $\sigma > 0$,
	\[
	\begin{aligned}
	\CRPS(F, y) &= y - 2\log(F(y)) - 1, \\
	\CRPS(F_{\mu, \sigma}, y) &= \sigma \CRPS\left(F, \tfrac{y - \mu}{\sigma} \right).
	\end{aligned}
	\]
	The CDFs are given by \(F_{\mu, \sigma}(x) = F\left(\tfrac{x - \mu}{\sigma}\right)\) and \(F(x) = \left(1 + \exp(-x)\right)^{-1}\).
	
	\subsubsection{Normal distribution}
	The function \fct{crps\char`_norm} computes the CRPS for the standard distribution, and generalizes via \code{mean} parameter $\mu \in \mathbb{R}$ and \code{sd} parameter $\sigma > 0$, or alternatively, \code{location} and \code{scale},
	\[
	\begin{aligned}
	\CRPS(\Phi, y) &= y\left(2\Phi(y)-1\right) + 2\varphi(y)  - \frac{1}{\sqrt{\pi}}, \\
	\CRPS(F_{\mu, \sigma}, y) &= \sigma \CRPS\left(\Phi, \tfrac{y - \mu}{\sigma} \right).
	\end{aligned}
	\]
	The CDFs are given by \(\Phi\) and \(F_{\mu, \sigma}(x) = \Phi\left(\tfrac{x - \mu}{\sigma}\right)\). Derived by \citet{GneitingEtAl2005}.
	
	\subsubsection{Mixture of normal distributions}	
	The function \fct{crps\char`_mixnorm} computes the CRPS for a mixture of normal distributions with mean parameters \(\mu_1, \ldots, \mu_M \in \mathbb{R}\) comprising \code{m}, scale parameters \(\sigma_1, \ldots, \sigma_M > 0\) comprising \code{s}, and (automatically rescaled) weight parameters \(\omega_1, \ldots, \omega_M > 0\) comprising \code{w},
	\[
	\CRPS(F,y) = \sum_{i=1}^M \omega_i A\left(y-\mu_i,\sigma_i^2\right) - \frac{1}{2} \sum_{i=1}^M\sum_{j=1}^{M}\omega_i \omega_j A\left(\mu_i-\mu_j,\sigma_i^2+\sigma_j^2\right).
	\]
	The CDF is \(F(x) = \sum_{i=1}^M \omega_i \Phi\left(\tfrac{x-\mu_i}{\sigma_i}\right)\), and \(A\left(\mu,\sigma^2\right) = \mu\left(2\Phi\left(\tfrac{\mu}{\sigma}\right) -1 \right) + 2\sigma \varphi\left(\tfrac{\mu}{\sigma}\right)\). Derived by \citet{GrimitEtAl2006}.
	
	\subsubsection[Student's t distribution]{Student's \(t\) distribution}
	The function \fct{crps\char`_t} computes the CRPS for Student's \(t\) distribution with \code{df} parameter \(\nu > 1\), and generalizes via \code{location} parameter \(\mu \in \mathbb{R}\) and \code{scale} parameter \(\sigma > 0\),
	\[
	\begin{aligned}
	\CRPS(F_\nu, y) &= y\Big(2F_\nu(y) - 1\Big) + 2f_\nu(y) \left(\frac{\nu + y^2}{\nu - 1}\right) - \frac{2\sqrt{\nu}}{\nu - 1}\frac{B(\tfrac{1}{2}, \nu - \tfrac{1}{2})}{B(\tfrac{1}{2}, \tfrac{\nu}{2})^2}, \\
	\CRPS(F_{\nu, \mu, \sigma}, y) &= \sigma\, \mathrm{CRPS}\left(F_\nu, \tfrac{y - \mu}{\sigma} \right).
	\end{aligned}
	\]
	The CDFs and PDF are given by \(F_{\nu, \mu, \sigma}(x) = F_\nu\left(\tfrac{x - \mu}{\sigma}\right)\) and
	\[
	\begin{aligned}
	F_\nu(x) &= \frac{1}{2} + \frac{x\ {}_2F_1(\tfrac{1}{2},\tfrac{\nu+1}{2};\tfrac{3}{2};-\tfrac{x^2}{\nu})}{\sqrt{\nu} B(\tfrac{1}{2},\tfrac{\nu}{2})}, \\
	f_\nu(x) &= \frac{1}{\sqrt{\nu}B(\tfrac{1}{2},\tfrac{\nu}{2})}\left(1+\frac{x^2}{\nu}\right)^{-\tfrac{\nu+1}{2}}. \\
	\end{aligned}
	\]
	
	\subsubsection{Two-piece exponential distribution}
	The function \fct{crps\char`_2pexp} computes the CRPS for the two-piece exponential distribution with \code{scale1} and \code{scale2} parameters \(\sigma_1, \sigma_2 > 0\), and generalizes via \code{location} parameter \(\mu \in \mathbb{R}\),
	\[
	\begin{aligned}
	\CRPS(F_{\sigma_1, \sigma_2}, y) &= \begin{cases}   \left\lvert y \right\rvert + \frac{2\sigma_1^2}{\sigma_1 + \sigma_2}\exp\left(-\left\lvert \frac{y}{\sigma_1}\right\rvert \right) - \frac{ 2\sigma_1^2}{\sigma_1 + \sigma_2} + \frac{\sigma_1^3 + \sigma_2^3}{2(\sigma_1 + \sigma_2)^2}, & y < 0, \\ \left\lvert y \right\rvert + \frac{2\sigma_2^2}{\sigma_1 + \sigma_2}\exp\left(-\left\lvert \frac{y}{\sigma_2}\right\rvert \right) - \frac{ 2\sigma_2^2}{\sigma_1 + \sigma_2} + \frac{\sigma_1^3 + \sigma_2^3}{2(\sigma_1 + \sigma_2)^2}, & y \ge 0, \end{cases} \\
    \CRPS(F_{\mu, \sigma_1, \sigma_2}, y) &= \mathrm{CRPS}(F_{\sigma_1, \sigma_2}, y - \mu).
	\end{aligned}
	\]
	The CDFs are given by \(F_{\mu, \sigma_1, \sigma_2}(x) = F_{\sigma_1, \sigma_2}(x - \mu)\) and
	\[
	F_{\sigma_1, \sigma_2}(x) = \begin{cases} \frac{\sigma_1}{\sigma_1 + \sigma_2}\exp\left(\frac{x}{\sigma_1}\right), & x < 0, \\
	1 - \frac{\sigma_2}{\sigma_1 + \sigma_2}\exp\left(-\frac{x}{\sigma_2}\right), & x \ge 0.\end{cases}
	\]
	
	\subsubsection{Two-piece normal distribution}
	The function \fct{crps\char`_2pnorm} computes the CRPS for the two-piece exponential distribution with \code{scale1} and \code{scale2} parameters \(\sigma_1, \sigma_2 > 0\), and generalizes via \code{location} parameter \(\mu \in \mathbb{R}\),
	\[
	\begin{aligned}
	\CRPS(F_{\sigma_1, \sigma_2}, y) &= \sigma_1\, \mathrm{CRPS}\left(F_{-\infty, 0}^{0, \sigma_2/(\sigma_1 + \sigma_2)}, \tfrac{\min(0, y)}{\sigma_1}\right) \\
	&\quad + \sigma_2\, \mathrm{CRPS}\left(F_{0, \sigma_1/(\sigma_1 + \sigma_2)}^{\infty, 0}, \tfrac{\max(0, y)}{\sigma_2}\right), \\
	\CRPS(F_{\mu, \sigma_1, \sigma_2}, y) &= \mathrm{CRPS}(F_{\sigma_1, \sigma_2}, y - \mu),
	\end{aligned}
	\]
	where \(F_{l, L}^{u, U}\) is the CDF of the generalized truncated/censored normal distribution as in Section~\ref{app:GenNormal}. The CDFs for the two-piece normal distribution are given by
	\[
	\begin{aligned}
	F_{\sigma_1,\sigma_2}(x) &= \begin{cases} \frac{2\sigma_1}{\sigma_1+\sigma_2}\Phi\left(\frac{x}{\sigma_1}\right), & x < 0,\\ \frac{\sigma_1-\sigma_2}{\sigma_1+\sigma_2} + \frac{2\sigma_2}{\sigma_1+\sigma_2} \Phi\left(\frac{x}{\sigma_2}\right), & x \ge 0, \end{cases} \\
	F_{\mu, \sigma_1, \sigma_2}(x) &= F_{\sigma_1, \sigma_2}(x - \mu).
	\end{aligned}
	\]
	\citet{GneitingThorarinsdottir2010} give an explicit CRPS formula.

	\subsection{Distributions for non-negative variables}
	
	\subsubsection{Exponential distribution}
	The function \fct{crps\char`_exp} computes the CRPS for the exponential distribution with \code{rate} parameter \(\lambda > 0\),
	\[
	\CRPS(F_\lambda, y) = |y| - \frac{2F_\lambda(y)}{\lambda} + \frac{1}{2\lambda}.
	\]
	The CDF is given by
	\[
	F_\lambda(x) = \begin{cases} 1 - \exp(-\lambda x), & x \ge 0, \\ 0, & x < 0.\end{cases}
	\]
	
	\subsubsection{Gamma distribution}
	The function \fct{crps\char`_gamma} computes the CRPS for the gamma distribution with \code{shape} parameter \(\alpha > 0\) and \code{rate} parameter \(\beta > 0\), or alternatively \code{scale = 1/rate},
	\[
	\CRPS(F_{\alpha,\beta},y) = y\left(2F_{\alpha,\beta}(y)-1\right) - \frac{\alpha}{\beta}\left(2F_{\alpha + 1, \beta}(y) -1\right) - \frac{1}{\beta B\left(\tfrac{1}{2},\alpha\right)}.
	\]
	The CDF is given by
	\[
	F_{\alpha,\beta}(x) = \begin{cases}\frac{\Gamma_l(\alpha,\beta x)}{\Gamma(\alpha)}, & x \geq 0, \\ 0, & x < 0. \end{cases}
	\]
	Derived by \citet{MoellerScheuerer2015}.
	
	\subsubsection{Log-Laplace distribution}
	The function \fct{crps\char`_llapl} computes the CRPS for the log-Laplace distribution with \code{locationlog} parameter \(\mu \in \mathbb{R}\) and \code{scalelog} parameter \(\sigma \in (0, 1)\),
	\[
	\CRPS(F_{\mu, \sigma}, y) = y\left(2F_{\mu, \sigma}(y) - 1\right) + \exp(\mu) \left(\tfrac{\sigma}{4 - \sigma^2} + A(y) \right).
	\]
	The CDF and otherwise required functions are given by
	\[\begin{aligned}
	F_{\mu, \sigma}(x) &= \begin{cases} 0, & x \leq 0,\\ \frac{1}{2}\exp\left(\frac{\log x - \mu}{\sigma}\right), & 0 < x < \exp(\mu), \\
	1 - \frac{1}{2}\exp\left(-\frac{\log x - \mu}{\sigma}\right), & x \geq \exp(\mu), \end{cases} \\
	A(x) &= \begin{cases} \frac{1}{1 + \sigma}\left(1-\left(2F_{\mu, \sigma}(x)\right)^{1+\sigma}\right), & x < \exp(\mu), \\ -\frac{1}{1-\sigma}\left(1-\left(2(1-F_{\mu, \sigma}(x))\right)^{1-\sigma}\right), & y \geq \exp(\mu). \end{cases}
	\end{aligned}\]
	
	\subsubsection{Log-logistic distribution}
	The function \fct{crps\char`_llogis} computes the CRPS for the log-logistic distribution with \code{locationlog} parameter \(\mu \in \mathbb{R}\) and \code{scalelog} parameter \(\sigma \in (0, 1)\),
	\[
	\begin{aligned}
	\CRPS(F_{\mu, \sigma}, y) &= y\left(2F_{\mu, \sigma}(y) - 1\right) \\
	&\quad - \exp(\mu)B(1 + \sigma, 1 - \sigma) \left(2\, I(1 + \sigma, 1 - \sigma, F_{\mu, \sigma}(y)) + \sigma - 1\right).
	\end{aligned}
	\]
	The CDF is given by
	\[
	F_{\mu, \sigma}(x) = \begin{cases} 0, & x \leq 0, \\ \left(1 + \exp\left(-\tfrac{\log x - \mu}{\sigma}\right)\right)^{-1}, & x > 0. \end{cases}
	\]
	\citet{TaillardatEtAl2016} give an alternative CRPS formula.
	
	\subsubsection{Log-normal distribution}
	The function \fct{crps\char`_lnorm} computes the CRPS for the log-logistic distribution with \code{locationlog} parameter \(\mu \in \mathbb{R}\) and \code{scalelog} parameter \(\sigma > 0\),
	\[
	\mathrm{CRPS}(F_{\mu,\sigma},y) = y\left(2F_{\mu, \sigma}(y) - 1\right) - 2 \exp(\mu+\sigma^2/2)\left(\Phi\left(\tfrac{\log y -\mu - \sigma^2}{\sigma}\right) + \Phi\left(\tfrac{\sigma}{\sqrt{2}}\right) - 1\right).
	\]
	The CDF is given by
	\[
	F_{\mu,\sigma}(x) = \begin{cases} 0, & x\leq 0, \\ \Phi\left(\tfrac{\log x - \mu}{\sigma}\right), & x > 0. \end{cases}
	\]
	Derived by \citet{BaranLerch2015}.
	
	\subsection{Distribution with flexible support and/or point masses}
	
	\subsubsection{Beta distribution}
	The function \fct{crps\char`_beta} computes the CRPS for the beta distribution with \code{shape1} and \code{shape2} parameters \(\alpha, \beta > 0\), and generalizes via \code{lower} and \code{upper} parameters \(l, u \in \mathbb{R}\), \(l < u\),
	\[
	\begin{aligned}
	\CRPS(F_{\alpha, \beta}, y) &= y(2F_{\alpha, \beta}(y) - 1) + \frac{\alpha}{\alpha + \beta} \left(1 - 2F_{\alpha + 1, \beta}(y) - \frac{2B(2\alpha, 2\beta)}{\alpha B(\alpha, \beta)^2} \right), \\
	\CRPS(F_{l, \alpha, \beta}^{u}, y) &= (u - l)\CRPS\left(F_{\alpha, \beta}, \tfrac{y - l}{u - l} \right).
	\end{aligned}
	\]
	The CDFs are given by \(F_{l, \alpha, \beta}^{u}(x) = F_{\alpha, \beta}\left(\tfrac{x - l}{u - l}\right)\) and
	\[
	F_{\alpha, \beta}(x) = \begin{cases} 0, & x < 0,\\ I(\alpha, \beta, x), & 0 \leq x < 1,\\ 1, & x \geq 1. \end{cases} \]
	\citet{TaillardatEtAl2016} give an equivalent expression.
	
	\subsubsection{Continuous uniform distribution}
	The function \fct{crps\char`_unif} computes the CRPS for the continuous uniform distribution on the unit interval, and generalizes via \code{min} and \code{max} parameters \(l, u \in \mathbb{R}\), \(l < u\), and by allowing point masses in the boundaries, i.e., \code{lmass} and \code{umass} parameters \(L, U \ge 0\), \(L + U < 1\),
	\[
	\begin{aligned}
	\CRPS(F, y) &= |y - F(y)| + F(y)^2 - F(y) + \frac{1}{3}, \\
	\CRPS(F_{L}^{U}, y) &= |y - F(y)| + F(y)^2(1 - L - U) - F(y)(1 - 2L) \\
	&\quad + \frac{(1 - L - U)^2}{3} + (1 - L)U, \\	
	\CRPS(F_{l, L}^{u, U}, y) &= (u - l) \CRPS\left(F_{L}^{U}, \tfrac{y - l}{u - l} \right).
	\end{aligned}
	\]
	The CDFs are given by \(F_{l, L}^{u, U}(x) = F_{L}^{U}\left(\tfrac{x - l}{u - l}\right)\) and
	\[
	\begin{aligned}
	F(x) &= \begin{cases} 0, & x < 0,\\ x, & 0 \leq x < 1,\\ 1, & x \ge 1, \end{cases} \\
	F_{L}^{U}(x) &= \begin{cases} 0, & x < 0, \\ L + (1 - L - U) x, & 0 \leq x < 1,\\ 1, & x \geq 1. \end{cases}
	\end{aligned}
	\]
	
	\subsubsection{Exponential distribution with point mass}
	The function \fct{crps\char`_expM} computes the CRPS for the standard exponential distribution, and generalizes via \code{location} parameter \(\mu \in \mathbb{R}\) and \code{scale} parameter \(\sigma > 0\), and by allowing a point mass in the boundary, i.e., a \code{mass} parameter \(M \in [0, 1]\),
	\[
	\begin{aligned}
	\CRPS(F_{M}, y) &= |y| - 2  (1 - M)F(y) + \frac{(1 - M)^2}{2}, \\
	\CRPS(F_{M, \mu, \sigma}, y) &= \sigma\, \mathrm{CRPS}\left(F_M, \tfrac{y - \mu}{\sigma} \right).
	\end{aligned}
	\]
	The CDFs are given by \(F_{M, \mu, \sigma}(x) = F_M\left(\tfrac{x - \mu}{\sigma}\right)\) and
	\[
	\begin{aligned}
	F(x) &= \begin{cases} 1 - \exp(-x), & x \ge 0, \\ 0, & x < 0, \end{cases} \\
	F_M(x) &= \begin{cases} M + (1 - M)F(x), & x \ge 0, \\ 0, & x < 0. \end{cases}
	\end{aligned}
	\]
	
	\subsubsection{Generalized extreme value distribution}
	The function \fct{crps\char`_gev} computes the CRPS for the generalized extreme value distribution with \code{shape} parameter \(\xi < 1\), and generalizes via \code{location} parameter \(\mu \in \mathbb{R}\) and \code{scale} parameter \(\sigma > 0\),
	\[
	\begin{aligned}
	\CRPS(F_\xi, y) &= \begin{cases} - y - 2 \operatorname{Ei}(\log F_{\xi}(y)) + \gamma - \log 2, & \xi = 0, \\
	y\left(2F_\xi(y) - 1\right) - 2G_\xi(y) - \frac{1 - \left(2 - 2^\xi\right)\Gamma(1 - \xi)}{\xi}, & \xi \neq 0, \end{cases} \\
	\CRPS(F_{\xi, \mu, \sigma}, y) &= \sigma\CRPS\left(F_{\xi}, \tfrac{y - \mu}{\sigma} \right).
	\end{aligned}\]
	The CDFs and otherwise required functions are given by \(F_{\xi, \mu, \sigma}(x) = F_{\xi}\left(\tfrac{x - \mu}{\sigma}\right)\) and
	\[
	\begin{aligned}	
	\text{for $\xi = 0$:}\quad F_{\xi}(x) &= \exp\left(-\exp(-x)\right) \\
	\text{for $\xi > 0$:}\quad F_{\xi}(x) &= \begin{cases} 0, & x \le -\frac{1}{\xi}, \\ \exp\left(-(1+\xi x)^{-1/\xi}\right), & x >  -\frac{1}{\xi}, \end{cases}\\
	G_{\xi}(x) &= \begin{cases} 0, & x \le -\frac{1}{\xi}, \\ -\frac{F_\xi(x)}{\xi} + \frac{\Gamma_u(1 - \xi, -\log F_\xi(x))}{\xi}, & x > -\frac{1}{\xi}, \end{cases} \\
	\text{for $\xi < 0$:}\quad F_{\xi}(x) &= \begin{cases} \exp\left(-(1+\xi x)^{-1/\xi}\right), & x <  -\frac{1}{\xi}, \\ 1, & x \ge -\frac{1}{\xi}, \end{cases} \\
	G_\xi(x) &= \begin{cases} -\frac{F_\xi(x)}{\xi} + \frac{\Gamma_u(1 - \xi, -\log F_\xi(x))}{\xi}, & x < -\frac{1}{\xi}, \\ -\frac{1}{\xi} + \frac{\Gamma(1 - \xi)}{\xi}, & x \ge -\frac{1}{\xi}. \end{cases}
	\end{aligned}
	\]
	\citet{FriederichsThorarinsdottir2012} give an equivalent expression.
	
	\subsubsection{Generalized Pareto distribution with point mass}
	The function \fct{crps\char`_gpd} computes the CRPS for the generalized extreme value distribution with \code{shape} parameter \(\xi < 1\), and generalizes via \code{location} parameter \(\mu \in \mathbb{R}\) and \code{scale} parameter \(\sigma > 0\), and by allowing a point mass in the lower boundary, i.e., a \code{mass} parameter \(M \in [0, 1]\),	
	\[
	\begin{aligned}
	\mathrm{CRPS}(F_{M, \xi}, y) &= |y| - \frac{2(1 - M)}{1 - \xi}\left(1 - \left(1 - F_\xi(y)\right)^{1 - \xi}\right) + \frac{(1 - M)^2}{2 - \xi}, \\
	\mathrm{CRPS}(F_{M, \xi, \mu, \sigma}, y) &= \sigma\, \mathrm{CRPS}\left(F_{M, \xi}, \tfrac{y - \mu}{\sigma} \right).
	\end{aligned}
	\]
	The CDFs are given by \(F_{M, \xi, \mu, \sigma}(x) = F_{M, \xi}\left(\tfrac{x - \mu}{\sigma}\right)\) and
	\[
	\begin{aligned}
	F_{M, \xi}(x) &= \begin{cases} M + (1 - M)F_\xi(x), & x \ge 0, \\ 0, & x < 0, \end{cases} \\
	\text{for $\xi = 0$:}\quad F_\xi(x) &= \begin{cases} 0, & x < 0, \\ 1 - \exp(-x), & x \ge 0, \end{cases} \\
	\text{for $\xi > 0$:}\quad F_\xi(x) &= \begin{cases} 0, & x < 0, \\ 1 - (1 + \xi x)^{-1/\xi}, & x \ge 0, \end{cases} \\
	\text{for $\xi < 0$:}\quad F_\xi(x) &= \begin{cases} 0, & x < 0, \\ 1 - (1 + \xi x)^{-1/\xi}, & 0 \le x < |\xi|^{-1}, \\ 1, & x \ge |\xi|^{-1}. \end{cases}
	\end{aligned}
	\]
	\citet{FriederichsThorarinsdottir2012} give a CRPS formula for the generalized Pareto distribution without a point mass.
	
	\subsubsection{Generalized truncated/censored logistic distribution}
	The function \fct{crps\char`_gtclogis} computes the CRPS for the generalized truncated/censored logistic distribution with \code{location} parameter \(\mu \in \mathbb{R}\), \code{scale} parameter \(\sigma > 0\), \code{lower} and \code{upper} boundary parameters \(l, u \in \mathbb{R}\), \(l < u\), and by allowing point masses in the boundaries, i.e., \code{lmass} and \code{umass} parameters \(L, U \ge 0\), \(L + U < 1\), 
	\[
	\begin{aligned}
	\CRPS\left(F_{l, L}^{u, U}, y\right) &= |y - z| + u U^2 - l L^2 \\
	&\quad - \left(\frac{1 - L - U}{F(u) - F(l)}\right)z \left(\frac{(1 - 2L)F(u) + (1 - 2U)F(l)}{1 - L - U}\right) \\
	&\quad - \left(\frac{1 - L - U}{F(u) - F(l)}\right)\left(2\log F(-z) - 2G(u)U - 2G(l)L\right) \\
	&\quad - \left(\frac{1 - L - U}{F(u) - F(l)}\right)^2 (H(u) - H(l)), \\
	\CRPS(F_{l, L, \mu, \sigma}^{u, U}, y) &= \sigma\CRPS\left(F_{(l - \mu)/\sigma, L}^{(u - \mu)/\sigma, U}, \tfrac{y - \mu}{\sigma} \right),
	\end{aligned}
	\]
	The CDFs are given by \(F(x) = \left(1 + \exp(-x)\right)^{-1}\) and
	\[
	\begin{aligned}
	F_{l, L}^{u, U}(x) &= \begin{cases} 0, & x < l, \\ \frac{1 - L - U}{F(u) - F(l)} F(x) - \frac{1 - L - U}{F(u) - F(l)} F(l) + L, & l \leq x < u, \\ 1, & x \geq u, \end{cases} \\
	F_{l, L, \mu, \sigma}^{u, U}(x) &= F_{(l - \mu)/\sigma, L}^{(u - \mu)/\sigma, U}\left(\tfrac{x - \mu}{\sigma}\right).
	\end{aligned}
	\]
	Otherwise required functions are given by \(G(x) = xF(x) + \log F(-x)\) and
	\[
	\begin{aligned}
	z &= \begin{cases} l, & y < l, \\ y, & l \le y < u, \\ u, & y \ge u, \end{cases} \\
	H(x) &= F(x) - xF(x)^2 + (1 - 2F(x))\log F(-x).
	\end{aligned}
	\]	
	The function \fct{crps\char`_clogis} computes the CRPS for the special case when the tail probabilities collapse into the respective boundary,
	\[
	\CRPS\left(F_{l}^{u}, y\right) = |y - z| +  z + \log \left(\frac{F(-l)F(u)}{F(z)^2}\right) - F(u) + F(l),
	\]
	where the CDF is given by
	\[
	F_{l}^{u}(x) = \begin{cases} 0, & x < l, \\ F(x), & l \leq x < u, \\ 1, & x \geq u. \end{cases}
	\]
	The function \fct{crps\char`_tlogis} computes the CRPS for the special case when \(L = U = 0\),
	where the CDF is given by
	\[
	F_{l}^{u}(x) = \begin{cases} 0, & x < l, \\ \frac{F(x) - F(l)}{F(u) - F(l)}, & l \leq x < u, \\ 1, & x \geq u. \end{cases}
	\]
	\citet{TaillardatEtAl2016} give a formula for left-censoring at zero. \citet{MoellerScheuerer2015} give a formula for left-truncating at zero.
	
	\subsubsection{Generalized truncated/censored normal distribution}\label{app:GenNormal}
	The function \fct{crps\char`_gtcnorm} computes the CRPS for the generalized truncated/censored normal distribution with \code{location} parameter \(\mu \in \mathbb{R}\), \code{scale} parameter \(\sigma > 0\), \code{lower} and \code{upper} boundary parameters \(l, u \in \mathbb{R}\), \(l < u\), and by allowing point masses in the boundaries, i.e., \code{lmass} and \code{umass} parameters \(L, U \ge 0\), \(L + U < 1\),
	\[
	\begin{aligned}
	\CRPS\left(F_{l, L}^{u, U}, y\right) &= |y - z| + u U^2 - l L^2 \\
	&\quad + \left(\frac{1 - L - U}{\Phi(u) - \Phi(l)}\right)z\left(2\Phi(z) - \frac{(1 - 2L)\Phi(u) + (1 - 2U)\Phi(l)}{1 - L - U}\right) \\
	&\quad + \left(\frac{1 - L - U}{\Phi(u) - \Phi(l)}\right)\left(2\varphi(z) - 2\varphi(u)U - 2\varphi(l)L\right) \\
	&\quad - \left(\frac{1 - L - U}{\Phi(u) - \Phi(l)}\right)^2 \left(\frac{1}{\sqrt{\pi}}\right) \left(\Phi\left(u\sqrt{2}\right) - \Phi\left(l\sqrt{2}\right)\right), \\
	\CRPS(F_{l, L, \mu, \sigma}^{u, U}, y) &= \sigma\CRPS\left(F_{(l - \mu)/\sigma, L}^{(u - \mu)/\sigma, U}, \tfrac{y - \mu}{\sigma} \right).
	\end{aligned}\]
	The CDFs and otherwise required functions are given by
	\[\begin{aligned}
	F_{l, L}^{u, U}(x) &= \begin{cases} 0, & x < l, \\ \frac{1 - L - U}{\Phi(u) - \Phi(l)} \Phi(x) - \frac{1 - L - U}{\Phi(u) - \Phi(l)} \Phi(l) + L, & l \leq x < u, \\ 1, & x \geq u, \end{cases} \\
	F_{l, L, \mu, \sigma}^{u, U}(x) &= F_{(l - \mu)/\sigma, L}^{(u - \mu)/\sigma, U}\left(\tfrac{x - \mu}{\sigma}\right), \\
	z&= \begin{cases} l, & y < l, \\ y, & l \le y < u, \\ u, & y \ge u. \end{cases}
	\end{aligned}\]
	The function \fct{crps\char`_cnorm} computes the CRPS for the special case when the tail probabilities collapse into the respective boundary,
	where the CDF is given by
	\[
	F_{l}^{u}(x) = \begin{cases} 0, & x < l, \\ \Phi(x), & l \leq x < u, \\ 1, & x \geq u. \end{cases}
	\]	
	The function \fct{crps\char`_tnorm} computes the CRPS for the special case when \(L = U = 0\),
	where the CDF is given by
	\[
	F_{l}^{u}(x) = \begin{cases} 0, & x < l, \\ \frac{F(x) - F(l)}{F(u) - F(l)}, & l \leq x < u, \\ 1, & x \geq u. \end{cases}
	\]
	\citet{ThorarinsdottirGneiting2010} give a formula for left-censoring at zero. \citet{GneitingEtAl2006} give a formula for left-truncating at zero.
	
	\subsubsection[Generalized truncated/censored Student's t distribution]{Generalized truncated/censored Student's \(t\) distribution}
	The function \fct{crps\char`_gtct} computes the CRPS for the generalized truncated/censored Student's \(t\) distribution with \code{df} parameter \(\nu > 1\), \code{location} parameter \(\mu \in \mathbb{R}\), \code{scale} parameter \(\sigma > 0\), \code{lower} and \code{upper} boundary parameters \(l, u \in \mathbb{R}\), \(l < u\), and by allowing point masses in the boundaries, i.e., \code{lmass} and \code{umass} parameters \(L, U \ge 0\), \(L + U < 1\),	
	\[
	\begin{aligned}
	\CRPS\left(F_{l, L, \nu}^{u, U}, y\right) &= |y - z| + u U^2 - l L^2 \\
	&\quad  + \left(\frac{1 - L - U}{F_\nu(u) - F_\nu(l)}\right) z\left(2F_\nu(z) - \frac{(1 - 2L)F_\nu(u) + (1 - 2U)F_\nu(l)}{1 - L - U}\right) \\
	&\quad - \left(\frac{1 - L - U}{F_\nu(u) - F_\nu(l)}\right)\left(2G_\nu(z) - 2G_\nu(u)U - 2G_\nu(l)L\right) \\
	&\quad - \left(\frac{1 - L - U}{F_\nu(u) - F_\nu(l)}\right)^2 \bar{B}_\nu \left(H_\nu(u) - H_\nu(l)\right), \\
	\CRPS(F_{l, L, \nu, \mu, \sigma}^{u, U}, y) &= \sigma \CRPS\left(F_{(l - \mu)/\sigma, L, \nu}^{(u - \mu)/\sigma, U}, \tfrac{y - \mu}{\sigma} \right).
	\end{aligned}
	\]
	The CDFs are given by
	\[
	\begin{aligned}
	F_\nu(x) &= \frac{1}{2} + \frac{x\ {}_2F_1\left(\tfrac{1}{2},\tfrac{\nu+1}{2};\tfrac{3}{2};-\tfrac{x^2}{\nu}\right)}{\sqrt{\nu} B\left(\tfrac{1}{2},\tfrac{\nu}{2}\right)}, \\
	F_{l, L, \nu}^{u, U}(x) &= \begin{cases} 0, & x < l, \\ \frac{1 - L - U}{F(u) - F(l)} F(z) - \frac{1 - L - U}{F(u) - F(l)} F(l) + L, & l \leq x < u, \\ 1, & x \geq u, \end{cases} \\
	F_{l, L, \nu, \mu, \sigma}^{u, U}(x) &= F_{\tfrac{l - \mu}{\sigma}, L, \nu}^{\tfrac{u - \mu}{\sigma}, U}\left(\tfrac{x - \mu}{\sigma}\right).
	\end{aligned}
	\]
	Otherwise required functions are given by
	\[
	\begin{aligned}
	z&= \begin{cases} l, & y < l, \\ y, & l \le y < u, \\ u, & y \ge u, \end{cases} \\
	f_\nu(x) &= \frac{1}{\sqrt{\nu}B\left(\tfrac{1}{2}, \tfrac{\nu}{2}\right)}\left(1 + \frac{x^2}{\nu}\right)^{-(\nu + 1)/2}, \\
	G_\nu(x) &= -\left(\frac{\nu + x^2}{\nu - 1}\right) f_\nu(x), \\
	H_\nu(x) &= \frac{1}{2} + \frac{1}{2}\, \mathrm{sgn}(x)\, I \left(\tfrac{1}{2}, \nu - \tfrac{1}{2}, \tfrac{x^2}{\nu + x^2}\right), \\
	\bar{B}_\nu &= \left(\frac{2\sqrt{\nu}}{\nu - 1}\right)\frac{B\left(\tfrac{1}{2}, \nu - \tfrac{1}{2}\right)}{B\left(\tfrac{1}{2}, \tfrac{\nu}{2}\right)^2}.
	\end{aligned}
	\]	
	The function \fct{crps\char`_ct} computes the CRPS for the special case when the tail probabilities collapse into the respective boundary,
	where the CDF is given by
	\[
	F_{l, \nu}^{u}(x) = \begin{cases} 0, & x < l, \\ F_\nu(x), & l \leq x < u, \\ 1, & x \geq u. \end{cases}
	\]	
	The function \fct{crps\char`_tt} computes the CRPS for the special case when \(L = U = 0\),
	where the CDF is given by
	\[
	F_{l, \nu}^{u}(x) = \begin{cases} 0, & x < l, \\ \frac{F_\nu(x) - F_\nu(l)}{F_\nu(u) - F_\nu(l)}, & l \leq x < u, \\ 1, & x \geq u, \end{cases}
	\]

	\subsection{Distribution for discrete variables}
	
	\subsubsection{Binomial distribution}
	The function \fct{crps\char`_binom} computes the CRPS for the binomial distribution with \code{size} parameter \(n = 0, 1, 2, \ldots,\), and \code{prob} parameter \(p \in [0, 1]\),
	\[
	\CRPS(F_{n, p}, y) = 2 \sum_{x = 0}^n f_{n, p}(x) \left(\ind\{y < x\} - F_{n, p}(x) + f_{n, p}(x)/2\right) (x - y).
	\]
	The CDF and probability mass function are given by
	\[
	\begin{aligned}
	F_{n, p}(x) &= \begin{cases} I\left(n - \lfloor x\rfloor, \lfloor x\rfloor + 1, 1 - p\right), & x \geq 0,\\
	0, & x < 0, \end{cases}\\
	f_{n, p}(x) &= \begin{cases} \binom{n}{x} p^x (1 - p)^{n - x}, & x = 0, 1, \ldots, n, \\ 0, & \text{otherwise}. \end{cases}
	\end{aligned}
	\]
	
	\subsubsection{Hypergeometric distribution}
	The function \fct{crps\char`_hyper} computes the CRPS for the hypergeometric distribution with two population parameters, the number \(m = 0, 1, \ldots,\) of entities with the relevant feature and the number \(n = 0, 1, \ldots,\) of entities without that feature, and a parameter for the size \(k = 0, \ldots, m + n\) of the sample to be drawn,
	\[
	\CRPS(F_{m, n, k}, y) = 2 \sum_{x = 0}^n f_{m, n, k}(x) \left(\ind\{y < x\} - F_{m, n, k}(x) + f_{m, n, k}(x)/2\right) (x - y).
	\]
	The CDF and probability mass function are given by
	\[
	\begin{aligned}
	F_{m, n, k}(x) &= \begin{cases} \sum_{i = 0}^{\lfloor x \rfloor} f_{m, n, k}(i), & x \geq 0,\\ 0, & x < 0,\end{cases} \\
	f_{m, n, k}(x) &= \begin{cases} \frac{\binom{m}{x} \binom{n}{k - x}}{\binom{m + n}{k}}, & x = \max\{0, k - n\}, \ldots, \min\{k, m\},\\ 0, & \text{otherwise}.\end{cases}
	\end{aligned}
	\]
	
	\subsubsection{Negative binomial distribution}
	The function \fct{crps\char`_nbinom} computes the CRPS for the negative binomial distribution with \code{size} parameter \(n > 0\), and \code{prob} parameter \(p \in (0, 1]\) or alternatively a non-negative mean parameter given to \code{mu},
	\[
	\begin{aligned}
	\CRPS(F_{n, p}, y) &= y\left(2F_{n, p}(y) - 1\right) \\
	&\quad - \frac{n(1-p)}{p^2}\left(p\left(2F_{n+1,p}(y-1) - 1\right) + {\ }_2F_1\left(n+1, \tfrac{1}{2}; 2; -\tfrac{4(1-p)}{p^2}\right)\right).
	\end{aligned}
	\]
	The CDF and probability mass function are given by
	\[
	\begin{aligned}
	F_{n, p}(x) &= \begin{cases} I\left(n, \lfloor x+1\rfloor, p\right), & x \geq 0, \\ 0, & x < 0, \end{cases} \\
	f_{n, p}(x) &= \begin{cases}\frac{\Gamma(x+n)}{\Gamma(n) x!} p^n (1-p)^x, & x = 0, 1, 2, \ldots, \\ 0, & \text{otherwise}. \end{cases}
	\end{aligned}
	\]
	Derived by \citet{WeiHeld2014}.
	
	\subsubsection{Poisson distribution}
	The function \fct{crps\char`_pois} computes the CRPS for the Poisson distribution with mean parameter \(\lambda > 0\) given to \code{lambda},
	\[
	\begin{aligned}
	\CRPS(F_\lambda, y) &= (y - \lambda) \left(2F_\lambda(y) - 1\right) + 2\lambda f_\lambda\left(\lfloor y\rfloor\right) - \lambda \exp(-2\lambda)\left(I_0(2\lambda) + I_1(2\lambda)\right).
	\end{aligned}
	\]
	The CDF and probability mass function are given by
	\[
	\begin{aligned}
	F_\lambda(x) &= \begin{cases} \frac{\Gamma_u(\lfloor x+1\rfloor, \lambda)}{\Gamma(\lfloor x+1 \rfloor)}, & x \geq 0,\\ 0, & x < 0, \end{cases}\\
	f_\lambda(x) &= \begin{cases}\frac{\lambda^x}{x!}e^{-\lambda}, & x = 0, 1, 2, \ldots, \\ 0, & \text{otherwise}, \end{cases}
	\end{aligned}
	\]
	Derived by \citet{WeiHeld2014}.

\section{Computation of multivariate scores for multiple forecast cases}\label{sec:multiv-multiple}

As noted in Section \ref{sec:multiv} the computation functions for multivariate scoring rules are defined for single forecast cases only. Here, we demonstrate how \code{apply} functions can be used to compute ES and VS$^p$ for multiple forecast cases. The simulation example is based on the function documentation of \fct{es\_sample} and \fct{vs\_sample}. 

The observation is generated as a sample from a multivariate normal distribution in $\R^{10}$ with mean vector $\boldsymbol{\mu} = (0,\dots,0)$ and covariance matrix $\boldsymbol{\Sigma}$ with $\boldsymbol{\Sigma}_{i,j} = 1$ if $i = j$ and $\boldsymbol{\Sigma}_{i,j} = c = 0.2$ if $i \neq j$ for all $i,j = 1,\dots,10$.
\begin{Schunk}
\begin{Sinput}
R> d <- 10
R> mu <- rep(0, d)
R> Sigma <- diag(d)
R> Sigma[!diag(d)] <- 0.2
\end{Sinput}
\end{Schunk}
The multivariate forecasts are given by 50 random samples from a corresponding multivariate normal distribution with mean vector $\boldsymbol{\mu}^f = (1,\dots,1)$ and covariance matrix ${\boldsymbol{\Sigma}}^f$ which is defined as $\boldsymbol{\Sigma}$, but with $c = 0.1$.
\begin{Schunk}
\begin{Sinput}
R> m <- 50
R> mu_f <- rep(1, d)
R> Sigma_f <- diag(d)
R> Sigma_f[!diag(d)] <- 0.1
\end{Sinput}
\end{Schunk}
The simulation process is independently repeated 1\,000 times. To illustrate two potential data structures, observations and forecasts are saved as list elements in an outer list where the index corresponds to the forecast case, and as 2- and 3-dimensional arrays where the last dimension indicates the forecast case.
\begin{Schunk}
\begin{Sinput}
R> n <- 1000
R> fc_obs_list <- vector("list", n)
R> obs_array <- matrix(NA, nrow = d, ncol = n)
R> fc_array <- array(NA, dim = c(d, m, n))
R> for (fc_case in 1:n) {
+    obs_tmp <- drop(mu + rnorm(d) %*% chol(Sigma))
+    fc_tmp <- replicate(m, drop(mu_f + rnorm(d) %*% chol(Sigma_f)))
+    fc_obs_list[[fc_case]] <- list(obs = obs_tmp, fc_sample = fc_tmp)
+    obs_array[, fc_case] <- obs_tmp
+    fc_array[, , fc_case] <- fc_tmp
+  }
\end{Sinput}
\end{Schunk}
Given the data structures of forecasts and observations, all 1\,000 forecast cases can be evaluated sequentially using the \fct{sapply} function (or, alternatively, a \code{for} loop) along the list elements or along the last array dimension.
\begin{Schunk}
\begin{Sinput}
R> es_vec_list <- sapply(fc_obs_list, function(x) es_sample(y = x$obs,
+    dat = x$fc_sample))
R> es_vec_array <- sapply(1:n, function(i) es_sample(y = obs_array[, i],
+    dat = fc_array[, , i]))
R> head(cbind(es_vec_list, es_vec_array))
\end{Sinput}
\begin{Soutput}
     es_vec_list es_vec_array
[1,]        2.44         2.44
[2,]        2.68         2.68
[3,]        2.56         2.56
[4,]        1.85         1.85
[5,]        3.83         3.83
[6,]        3.04         3.04
\end{Soutput}
\end{Schunk}

\end{appendix}

\end{document}